\documentclass[preprint,5p,number]{elsarticle}

\usepackage{amssymb}
\usepackage{amsmath}
\usepackage{amsthm}
\theoremstyle{definition}
\newtheorem{definition}{Definition}
\usepackage{graphicx}
\usepackage{wrapfig}
\usepackage{threeparttable}
\usepackage{longtable}
\usepackage{supertabular}
\usepackage{amssymb}
\usepackage{xcolor}
\usepackage{microtype}
\usepackage{capt-of}

\usepackage{listings}
\lstdefinelanguage{rebeca}{
    morekeywords={reactiveclass, knownrebecs, statevars, main, msgsrv, constraints,con, main, define, LTL, CTL, wait, msg, priority, reset, set, self, now, after, delay, deadline, initial, unicast, succ, unsucc, Priority, true, false, int, byte, boolean, else, if},
	otherkeywords={=>,<-,<\%,<:,>:,\#,@},
	sensitive=true,
	morecomment=[l]{//},
	morecomment=[n]{/*}{*/},
	morestring=[b]",
	morestring=[b]',
	morestring=[b]"""
}

\usepackage{cleveref}
\usepackage{tabularx}
\usepackage{array}
\usepackage{algorithm}
\usepackage{algpseudocode} 
\algnewcommand{\algorithmicswitch}{\textbf{switch}}
\algnewcommand{\algorithmicendswitch}{\textbf{end switch}}
\algnewcommand{\algorithmiccase}{\textbf{case}}
\algnewcommand{\algorithmicdefault}{\textbf{default}}
\algnewcommand{\Switch}[1]{\State \algorithmicswitch\ #1}
\algnewcommand{\EndSwitch}{\State \algorithmicendswitch}
\algnewcommand{\Case}[1]{\State \hspace{\algorithmicindent}\algorithmiccase\ #1:}
\algnewcommand{\Default}{\State \hspace{\algorithmicindent}\algorithmicdefault:}
\algrenewcommand\algorithmicfunction{\textbf{method}}
\usepackage{caption}
\newcounter{requirements}
\newcounter{properties}
\newcolumntype{L}{>{\raggedright\arraybackslash}X}
\newcolumntype{R}{>{\raggedleft\arraybackslash}X}
\newcolumntype{C}{>{\centering\arraybackslash}X}

\newcommand{\Requirement}[1]{%
  \par\addvspace{\topsep}% <----- Adjust to suit
  \noindent\quad % <------------- Are you sure about this?
  \refstepcounter{requirements}%
  \textbf{Requirement R\arabic{requirements}:}%
  \quad
  #1%
  \par
  \addvspace{\topsep}% <--------- Adjust to suit
}

\newcommand{\Property}[1]{%
  \par\addvspace{\topsep}% <----- Adjust to suit
  \noindent % <------------- Are you sure about this?
  \refstepcounter{properties}%
  \textbf{Property P\arabic{properties}:}%  
  #1%
  \par
  \addvspace{\topsep}% <--------- Adjust to suit
}

\journal{Computers and Security}
\begin{document}

\begin{frontmatter}

%% Title, authors and addresses

%% use the tnoteref command within \title for footnotes;
%% use the tnotetext command for theassociated footnote;
%% use the fnref command within \author or \affiliation for footnotes;
%% use the fntext command for theassociated footnote;
%% use the corref command within \author for corresponding author footnotes;
%% use the cortext command for theassociated footnote;
%% use the ead command for the email address,
%% and the form \ead[url] for the home page:
%% \title{Title\tnoteref{label1}}
%% \tnotetext[label1]{}
%% \author{Name\corref{cor1}\fnref{label2}}
%% \ead{email address}
%% \ead[url]{home page}
%% \fntext[label2]{}
%% \cortext[cor1]{}
%% \affiliation{organization={},
%%            addressline={}, 
%%            city={},
%%            postcode={}, 
%%            state={},
%%            country={}}
%% \fntext[label3]{}

\title{From Automata Learning to Model Checking: Formal Security Verification of Black-Box Protocols} %% Article title

%% use optional labels to link authors explicitly to addresses:
%% \author[label1,label2]{}
%% \affiliation[label1]{organization={},
%%             addressline={},
%%             city={},
%%             postcode={},
%%             state={},
%%             country={}}
%%
%% \affiliation[label2]{organization={},
%%             addressline={},
%%             city={},
%%             postcode={},
%%             state={},
%%             country={}}

\author[MDU,AVL,AIT]{Stefan Marksteiner\corref{mycorrespondingauthor}}
\cortext[mycorrespondingauthor]{Corresponding author. M\"{a}lardalen University, 721 23 V\"{a}ster\r{a}s, , Box 883, Sweden.}
\ead{stefan.marksteiner@mdu.se}
\author[MDU]{Mikael Sj\"{o}din}
\author[MDU,KTH]{Marjan Sirjani}

\address[MDU]{M\"{a}lardalen University, V\"{a}ster\r{a}s, Sweden}
\address[AVL]{AVL List GmbH, Graz, Austria}
\address[AIT]{AIT Austrian Institute of Technology GmbH, Graz, Austria}
\address[KTH]{KTH Royal Institute of Technology, Stockholm, Sweden}

%% Author affiliation
%\affiliation{organization={},%Department and Organization
%            addressline={}, 
%            city={},
%            postcode={}, 
%            state={},
%            country={}}

%% Abstract
\begin{abstract}
%% Text of abstract
Security verification of communication protocols in industrial and safety-critical systems is challenging because implementations are often proprietary, accessible only as black boxes, and too complex for manual modeling. As a result, existing security testing approaches usually depend on incomplete test suites and/or require labor-intensive modeling, limiting coverage, scalability, and trust. This paper addresses the problem of systematically verifying protocol security properties without access to internal system models.
We propose a flexible and scalable method for formal verification of communication protocols that combines active automata learning with model checking to enable rigorous security analysis of black-box protocol implementations. Behavioral models are first inferred from system interactions using automata learning.
We then propose context-based proposition maps (CPMs) to enrich the learned models with semantic information, yielding annotated Mealy machines that bridge the gap between learned behavior and property verification. These machines are automatically transformed into verifiable models in the Rebeca modeling language, enabling model checking of generic security properties such as authentication, confidentiality, privilege levels, and key validity, while allowing protocol-specific properties to be manually added.
The CPMs are also used to populate these generic properties with protocol-specific propositions. Furthermore, the resulting model can be easily altered to introduce non-deterministic behavior (like timeouts or faults) and examined if the properties still hold under these different conditions. As a result, we gain an automated tool chain that spans from model learning to security property checking of specific protocols.
The key contributions include: (i) a method for augmenting learned protocol models with security-relevant propositions, (ii) a fully automated transformation pipeline from learned models to model-checking artifacts, (iii) reusable, generic security property templates that are instantiated in protocol-specific models, and (iv) empirical validation through case studies demonstrating applicability in different protocols and domains. The results show that the approach enables scalable and systematic discovery of security vulnerabilities in black-box systems while reducing modeling effort and improving automation compared with traditional verification workflows.

\end{abstract}

%%Graphical abstract
%\begin{graphicalabstract}
%\includegraphics{grabs}
%\end{graphicalabstract}

%%Research highlights
%\begin{highlights}
%\item Research highlight 1
%\item Research highlight 2
%\end{highlights}

%% Keywords
\begin{keyword}
%% keywords here, in the form: keyword \sep keyword
Automata Learning \sep Model Checking \sep Cybersecurity \sep NFC \sep UDS \sep Rebeca \sep Formal Methods
%% PACS codes here, in the form: \PACS code \sep code

%% MSC codes here, in the form: \MSC code \sep code
%% or \MSC[2008] code \sep code (2000 is the default)

\end{keyword}

\end{frontmatter}

%\linenumbers

\section{Introduction}

%\subsection{Motivation}
Interconnected electronic systems in all domains of everyday life are becoming increasingly ubiquitous and complex. Various domains like automotive, manufacturing, but also critical infrastructures like electronic government systems naturally have to provide communication capabilities to provide smarter and more usable functionality. It is therefore crucial to create methodologies to verify the correctness and security of these systems. Often those systems are proprietary, which means that verifying their security has become increasingly laborious and uncertain. Many approaches rely on static, pre-defined test suites that do not take any architectural or behavioral information into account. 
Therefore, validation of industrial systems and critical infrastructure often include the use of formal methods that provide more rigor in verification~\cite{terbeekFormalMethodsIndustry2024}. Such methods, including model checking, often come with the disadvantage of a high effort (especially with modeling systems) in conjunction with being hard to apply at black-box systems. To address these disadvantages, we utilize models from active automata learning, which is integrated in our tool chain~\cite{marksteiner_learning_2025}. This technique allows for an automated model generation to be used in a checking process, and is also a technique that is suitable for black-box systems~\cite{vaandragerModelLearning2017}. 
Combining these two techniques of formal methods, we create a tool chain for checking interactive systems (particularly their communication protocols) for  correctness and security.
To achieve this, we make automatically learned models applicable to model checking and by providing generic security properties that can easily be instantiated for concrete protocols. The core idea for both are \textit{context-based proposition maps} (\textit{CPMs}). CPMs are simple, tabular rule sets 
that are used to define when certain propositions become true or false by actions on the transitions from one state to another. For the modeling, we use CPMs to annotate Mealy machines~\cite{mealy_method_1955}, obtained via automata learning, with checkable propositions.
 The result of the annotation process are hybrid structures between Mealy machines and Kripke structures we call \textit{annotated Mealy machines (AMMs)}. Creating a formally defined template, we also provide an automated translation of AMMs into the Rebeca modeling language~\cite{sirjaniRebecaTheoryApplications2006,sirjani_modeling_2004}, where it can be run through a model checker.
For the generic property definition (particularly authentication, confidentiality, privilege levels, and key validity), we use composition of  propositions that we also define through CPMs. This makes the properties universally applicable, as long as a CPM is defined. 
The motivation to perform each step is as follows:
\begin{itemize}
%To be able to formally verify we need a model of the system. 
    \item To be able to formally verify a system, we need a suitable model thereof that contains propositions we can use to formulate meaningful properties to check
%Building models manually is not straightforward, so, for gaining the models we use AAL. 
% Formal Verification needs model, so AAL ..
    \item As building models of complex systems manually is time-intensive and error-prone, we use automata learning to infer a model from the system
%AAL-derived models only captures the change of states and the actions that change the state but the states are abstract. 
% Automata lack meaning, so, context
    \item Learned models are, however, abstract and only capture the change of states and the actions that change the state, but not the context needed to form meaningful propositions
%%We still lack the context, the meaningful propositions which we need to check in formal verification. (Algoritmic mapping of the actions to some propositions and put them in the states is not enough.) 
%We need to connect the "required attributes" to state propositions to give us the "context".
% In this paper, we propose an approach to automate adding the "context" that works for both making the model more concrete and properties to be checked.
%We introduce CPMs to add context to the models, so we can check them.
    \item We therefore propose an approach to automatically annotate learned models and generic properties with propositions according to a lightweight table-based rule set (CPMs)
    \item Lastly, we transform the annotated model into a modeling language (in our case Rebeca) that is expressive enough to be able to both perform model checking and to alter the model to simulate hypothetical properties (e.g., introducing potential new attack vectors, security controls, and non-deterministic behavior)
\end{itemize}
If the protocol is known, CPMs are easy to create. The only task is to denote which actions in a protocol changes a proposition to true or false. This action is represented by an input or output as a label on a transition. We call this change of value of a proposition as gain or loss of the proposition in a state when the propositions becomes true or false, respectively.
In summary, our main contributions are: 
\begin{itemize}
    \item 
    Proposing \textit{context-based proposition maps (\textit{CPMs}) to be used for annotating the generic models and populating the generic properties with protocol-specific propositions,}
    
    \item 
    Offering an approach for automatically annotating learned models (Mealy machines) with propositions using CPMs and
    creating \textit{annotated Mealy machines (AMMs)} -- hybrid structures of Mealy machines and Kripke structures (suitable for model checking),
    
    \item Defining and implementing a
    template that allows for fully automatically creating Rebeca models from AMMs,
    
    \item Defining generalized properties to check security requirements along with an easily usable and scalable method for 
    deriving concrete properties based on the  generalized ones using CPMs,
    
    \item Demonstrating the applicability and versatility of the approach using two case studies of communication protocols (NFC~\cite{internationalorganizationforstandardizationCardsSecurityDevices2018a} and UDS~\cite{internationalorganizationforstandardizationRoadVehiclesUnified2020}).
\end{itemize}

The remainder of this paper outlines as follows: Section \ref{sec:pre} contains basics. Section \ref{sec:approach} outlines the approach, Section \ref{sec:check:AP:CPM} the concept of context-based proposition maps, Section \ref{sec:check:AP} the process of annotating Mealy machines with propositions, Section \ref{sec:gen} the Rebeca code generation including a formally defined template and a formal description of the CPM concept. Section \ref{sec:check} the model checking, including property definition. Section \ref{sec:test} briefly describes the relation to testing. Section \ref{sec:eval} presents the case studies, Section \ref{sec:rel} the related work and Section \ref{sec:conc} concludes the paper.

\section{Background}\label{sec:pre}
In this section, we briefly describe the basics of automata, automata learning, model checking, as well as the frameworks we used and protocols used in the case studies.

\subsection{Automata and Kripke Structures}
\label{sec:pre:sm}
Automata, or state machines, are a fundamental concept of computer science
~\cite{turingComputableNumbersApplication1937,neumannGeneralLogicalTheory1968}.
To describe reactive systems, we use a certain type of automata, namely Mealy machines~\cite{mealy_method_1955}. A Mealy machine $\mathcal{M}$ is defined as $\mathcal{M}=(Q,\Sigma,\Omega,\delta,\lambda,q_0)$, with $Q$ being the set of states, $\Sigma$  the input alphabet, $\Omega$ the output alphabet (that may or may not be identical to the input 
alphabet), $\delta$ the transition function ($\delta: Q \times \Sigma \rightarrow Q$), $\lambda$ the output function ($\lambda: Q \times \Sigma \rightarrow \Omega$), and $q_0$ the initial state. An automaton can describe a system such that for any defined set of inputs (input word), it delivers the correct set of outputs (output word)~\cite{rabinFiniteAutomataTheir1959}.
While Mealy machines are widely used to model system behavior, Kripke structures~\cite{kripke_semantical_1963} are basic structures used for model checking~\cite{clarke_birth_2008}. A Kripke structure is similar to a Mealy machine, as both can be seen as types of Labeled Transition Systems (LTS)~\cite{tapplerModelBasedTestingIoT2017,clarke_model_1999}. However, they differ in that instead of labeling the transitions, Kripke structures have their states labeled with \textit{propositions}. Coming from propositional logic, these are logical attributes that can be true or false. Any state in which this attribute is true is labeled with that proposition. Each state can have an element of the power set of all propositions. The formal definition of a Kripke structure ($\mathcal{K}$) is $\mathcal{K}= (S,S_0,R, \mathcal{L})$ with $S$ being the set of states, $S_0 \subseteq S$ the set of input states, $R \subseteq S \times S$ a transition relation and $\mathcal{L}= S \rightarrow 2^{P}$ the labeling function that attributes an element of the power set of (atomic) propositions (${p \in P}$) to the states.

\subsection{Linear Temporal Logic and Model Checking}
\label{sec:back:LTL}
Linear Temporal Logic (LTL) is an extension of propositional logic~\cite{russellMathematicalLogicBased1908} that allows for expressing logical temporal modalities~\cite{baierPrinciplesModelChecking2008}. In this context, \textit{temporal} is not to be confused with \textit{timed}. LTL only allows statements about the modality and succession order of events to occur, not about a duration of any kind. That is, LTL combines atomic propositions (attributes of a state that can be true or false) formulas using logical operators and extends them with the following modalities: 
\begin{itemize}
    \item \textit{always} ($\square$, or $G$ for Globally): the proposition must hold in any subsequent state
    \item \textit{eventually} ($\lozenge$, or $F$ for Finally): the proposition must hold in some (i.e., any arbitrary) subsequent state (may or may not hold before)
    \item \textit{next} ($\bigcirc$, or $X$ for neXt): the proposition must hold in the immediately subsequent state and
    \item \textit{until} ($\mathcal{U}$ or $U$): the proposition $A_1$ must hold until another defined proposition $A_2$ occurs ($A_1\mathcal{U}A_2$). $A_1$ may or may not hold \textit{after} $A_2$ has occurred. In any case, $A_2$ has to occur at some point.
\end{itemize}
A model checker runs through all states of a model and checks if their properties hold~\cite{baierPrinciplesModelChecking2008}. The properties are often expressed in LTL or related logics like Computation tree logic (CTL)~\cite{pnueliTemporalLogicPrograms1977} or branching-time logic (CTL*)~\cite{allen_emerson_deciding_1984}. In automata-theoretic model checking, a model checker ordinarily creates B{\"u}chi automata of both the scrutinized model and the negation of the LTL property to check, and examines if their cross product's accepted language is empty~\cite{peled_black_1999}.

\subsection{Rebeca}
\label{sec:intro:check:reb}
Rebeca is a modeling language with model checking support specifically designed to model and model check reactive (i.e., communicating) systems~\cite{sirjaniRebecaTheoryApplications2006}. Rebeca resembles the Java programming language to ease its usage for engineers, and is used in several applications~\cite{DBLP:conf/birthday/2025sirjani}. It conceptually uses \textit{Rebecs} to model reactive actors. Rebecs are similar to Java objects,
but possess interfaces (\textit{message servers}) to interact with each other and contain a message queue to process messages from other Rebecs in a FIFO manner. 
Its IDE, Afra~\cite{khamespanahAfraEclipseBasedTool2023}, contains a model checking framework based on  Modere~\cite{jaghooriModereModelcheckingEngine2006}. 
Each Rebec can have state variables that can constitute checkable properties. The properties are defined as LTL formula in a separate  property file.
%Apart from the usability for software engineering and modeling and model checking reactive systems, it is possible to convert code into Lingua Franca~\cite{sirjaniModelCheckingSoftware2020}. Lingua Franca is a coordination language, providing a deterministic concurrency model, that supports C, C++, Python, TypeScript, and Rust as targets~\cite{menardHighperformanceDeterministicConcurrency2023}. Eventually, a Rebeca code can produce a verified program in one of these languages via Lingua Franca. 
Formally, a Rebeca model $\mathcal{R}$ is defined as $\mathcal{R}=||_{i\in\mathcal{I}}\text{ }r_i$, which is the set of concurrent actors (also called \textit{rebec} for \textit{re}active o\textit{b}j\textit{ec}t) $r_i$, with $i \in \mathcal{I}$ being an identifier in the set of all identifiers in the model~\cite{sirjani_modeling_2004}. An actor is defined as $r_i:=\langle V_i,M_i,K_i\rangle$, where $i$ is a unique identifier, $Vi$ the set of state variables, $M_i$ the set of method identifiers (local methods and remotely callable methods called \textit{message servers}), and $K_i$ the list of other actors known to actor $i$. Remote method (i.e., message server) calls between actors occur by sending messages ${msg}=\langle{sendid},i,{mtdid}\rangle$, with ${sendid} \in \mathcal{I}$ being the sender actor's identifier, $i \in \mathcal{I}$ the receiver actor's identifier, and $mtdid \in M_i$ the identifier of the method called on the sender side.

 \begin{figure}[t]%[ht!]
	    \includegraphics[width=\linewidth]{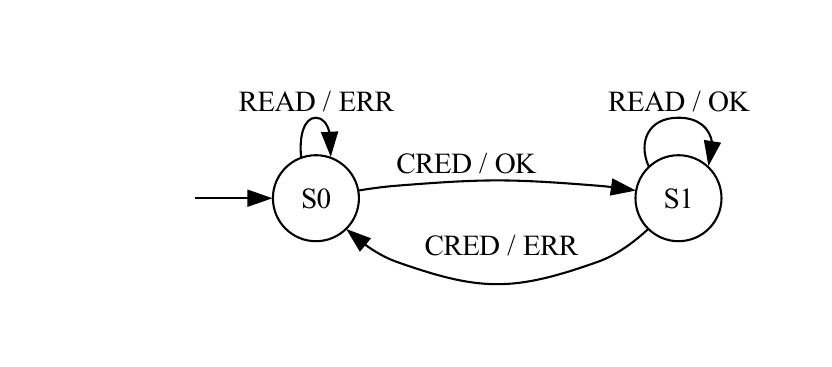}
\caption{A graphical Mealy machine representation of an example for a simple authenticated file read mechanism.}
\label{fig:ExampleMealy}
\end{figure}

\section{Approach}
\label{sec:approach}
\begin{figure*}[t]%[ht!]
	    \includegraphics[width=\linewidth]{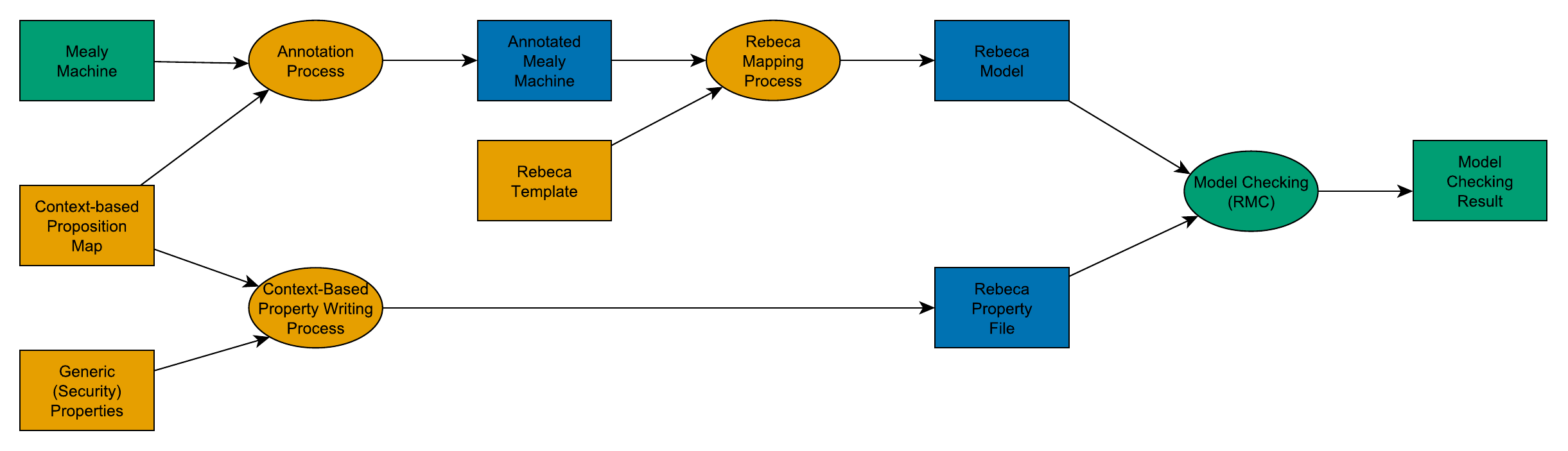}
\caption{Overview of the proposed approach; boxes denote artifacts and ellipses denote processes or tools. Orange items defined in this paper, blue are automatically generated artifacts, and green are used items from other works.}
\label{fig:Process}
\end{figure*}

%We therefore follow the approach to translate the learned models into a modeling language, Rebeca~\cite{sirjaniRebecaTheoryApplications2006}. 
We utilize active automata learning using the \textit{LearnLib} framework to infer Mealy machine models of a respective System-under-learning (SUL). 
%We therefore created dedicated protocol adapters and defined input alphabets (Section \ref{sec:learner}). 
The details of the learning process are out of scope of this paper but we describe our learning tool chain in another paper~\cite{marksteiner_learning_2025}.
Since this model does not intrinsically contain any properties, we annotate it with sensible propositions, creating a  hybrid automaton of a Mealy machine and a Kripke structure, which we call an \textit{annotated Mealy machine (AMM)}.
For that, we define \textit{context-based proposition maps (CPMs)} that determine the rule set when propositions become true or false (Section \ref{sec:check:AP:CPM}). We created a tool that uses these maps to annotate Mealy machines represented in the GraphViz format, receiving an AMM. We then  convert this structure into Rebeca code and verify it to be faithful to the original model in an automated way, formally defining a Rebeca template for that purpose (Sections \ref{sec:gen} and \ref{sec:check:verify}) . Once this step has been completed, we utilize the Rebeca Model Checking tool, RMC~\cite{khamespanahAfraEclipseBasedTool2023} to check the models for properties defined in linear temporal logic (LTL). These properties are partly specific to the scrutinized protocol, but can also partly be generalized to check very diverse protocols 
for the same properties:  we define generic, commonly important, properties like authentication, confidentiality, privilege levels, and key validity (Section \ref{sec:check:rules}), while each protocol might still have  constraints that are unique to it (like the order of security-related commands).
Finally, if a property violation is found, the trace can be concretized via the learner's interface and be used as a test case for verification on the live system. If this confirms the violation, that indicates an issue. Otherwise, the different behavior serves as a counterexample for the learner in a new iteration of the process. 
That way, we implemented all the phases for going through learning, checking, and testing real-world examples. 
We automated the complete process by combining (a) the \textit{learner(s)} using LearnLib~\cite{isberner_opensource_2015}, 
(b) a tool that generates an \textit{annotated Mealy machine} and translating it into a Rebeca model to be checked (Section \ref{sec:gen}) and (c) the Rebeca \textit{model checking} tool 
\cite{khamespanahAfraEclipseBasedTool2023}, used with generic security properties. We emphasize that this work concentrates on part b), providing the link between automata learning (see~\cite{marksteiner_learning_2025} for details of our tool chain for obtaining models) and model checking.
Figure \ref{fig:Process} provides an overview of the proposed automated process of model checking learned Mealy machines for generic security properties, which basically consists of two streams: the model annotation that transfers the learned Mealy machines into a checkable format and the context-based property writing that delivers the checking rules. The former takes a Mealy machine (as described, we learn it automatically) and a CPM as an input, which an algorithm turns into an AMM. Using our Rebeca template, another algorithm further transforms the AMM into a Rebeca model. The second stream takes the generic properties and the same CPM as an input an uses an algorithm to define the composing propositions according to the CPM and writes a Rebeca property file. All three mentioned algorithms are implemented in a Java class that takes the Mealy machine, the CPM and generic properties (we also define later in this paper) as inputs and generates a Rebeca model and property file as outputs. We feed these two outputs Rebeca Model Checker (RMC) that automatically checks the resulting model for the resulting security properties.

\section{Context-Based Proposition Maps}
\label{sec:check:AP:CPM}
Once we obtain a Mealy machine of the SUL (in GraphViz format), we create Context-Based Proposition Maps (CPMs) to annotate the model with propositions of our interest.
%The generic requirements we defined in this paper are the basis for an engineer to populate the CPM table.
%We defined generic properties which constitute of generic propositions like authentication and successful access. The engineer makes these generic propositions specific in the CPM table  based on the specific protocol. 
Each state in the Mealy machine is augmented with  propositions which will also be included in the Rebeca model and will be used for checking the desired properties while model checking.  

\subsection{Motivating Example}
\label{sec:check:AP:CPM:Example}
To give an illustrative example, we define a Mealy machine that represents a trivial authenticated file access. 
The machine has two states ($Q=\{S0,S1\}$), two input ($\Sigma=\{$CRED,\allowbreak READ$\}$), and two output ($\Omega=\{$OK,\allowbreak ERR$\}$) symbols. Further, we have the initial state $q_0=S0$,
the transition function 
\\$\delta(q, \sigma) :=
\begin{cases}
S0 & \text{ if } q = S0 \land \sigma = \text{ READ}\\
&|| \text{ } q = S1 \land \sigma = \text{ CRED};\\
S1 & \text{ if } q = S0 \land \sigma = \text{ CRED}\\
&|| \text{ } q = S1 \land \sigma = \text{ READ};
\end{cases}$
\\and the output function 
\\$\lambda(q, \sigma) := 
\begin{cases}
\text{OK} & \text{if } q = S0 \land \sigma = \text{ CRED}\\
&|| \text{ } q = S1 \land \sigma = \text{ READ};\\
\text{ERR} & \text{if } q = S0 \land \sigma = \text{ READ}\\
 &|| \text{ } q = S1 \land \sigma = \text{ CRED;}
\end{cases}$

\noindent
Figure \ref{fig:ExampleMealy} shows a graphical representation of this example. This machine essentially describes that before reading, users must provide their credentials ($CRED$) to authenticate and that providing credentials twice de-authenticates. We will re-use this example Mealy machine to demonstrate the usage of CPMs to annotate Mealy machines with propositions and to produce Rebeca code.

\subsection{Propositions and Mealy Machines}
Automata learning provides us with a SUL's Mealy machine just like our example.
In a Mealy machine we only have inputs and outputs to distinguish different states and each state does not include any other information.
Our goal is to determine the security of a system based on certain properties and it is more intuitive to specify the properties based on the information about the states. We therefore need to annotate the states with relevant propositions.
These propositions should provide the necessary 
context for sensible security checking. 
We define intuitive and usable propositions 
for our context for each state (e.g., authenticated) and provide a method to annotate the 
model with these propositions in a correct way. Since the only way of determining the propositions  is through the system's behavior (i.e., input and output), a straightforward approach is to define a map to show which input and output would make a 
proposition true or false. We call this map a \textit{Context-based Proposition Map (CPM)}.
Each row in the CPM table  defines a \textit{condition}. Table \ref{tab:CPMex} shows a CPM that we use to annotate the example Mealy machine from Section \ref{sec:check:AP:CPM:Example} with the propositions $AUTH$, $PROT$, and $ACCESSOK$ in a semantically sensible way with regard to the protocol. Later, in the model checking process, these propositions form security properties we want to check.

\begin{table}[t!]
\caption{CPM for the Example Mealy.}
\label{tab:CPMex}
\begin{threeparttable}

\begin{tabularx}{\columnwidth}{>{\centering\arraybackslash}X |
                                >{\centering\arraybackslash}X |
                                >{\centering\arraybackslash}X}
\multicolumn{3}{c}{}\\
\hline
\textbf{Proposition} & \textbf{Input} & \textbf{Output} \\
\hline
\multicolumn{3}{c}{Gains ($C_g$)} \\
\hline
AUTH & CRED & OK \\ 
PROT & *    & *  \\ 
\hline
\multicolumn{3}{c}{Losses ($C_l$)} \\
\hline
AUTH & CRED & ERR \\ 
\hline
\multicolumn{3}{c}{Implicit State Propositions ($C_\tau$)} \\
\hline
ACCESSOK & READ & OK \\
\hline
\end{tabularx}

\end{threeparttable}
\end{table}

\begin{figure*}[ht!]%[ht!]
	    a)\includegraphics[width=0.19\linewidth]{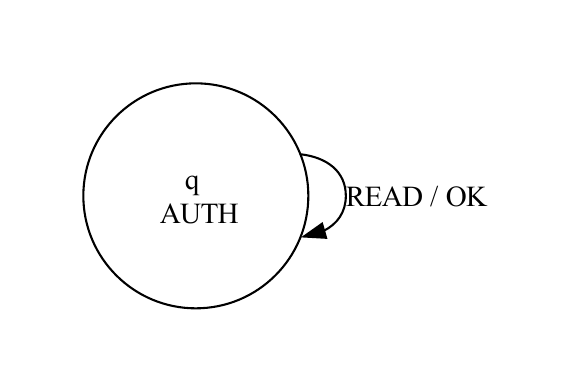}
        \includegraphics[width=0.29\linewidth]{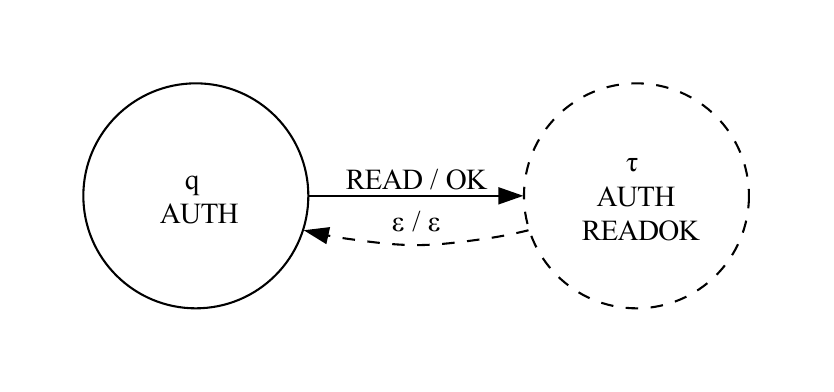}
        b)\includegraphics[width=0.19\linewidth]{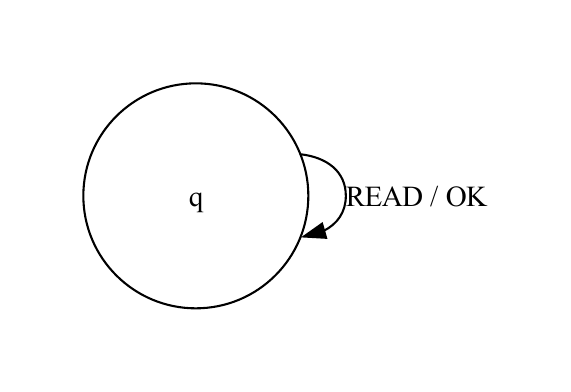}
        \includegraphics[width=0.29\linewidth]{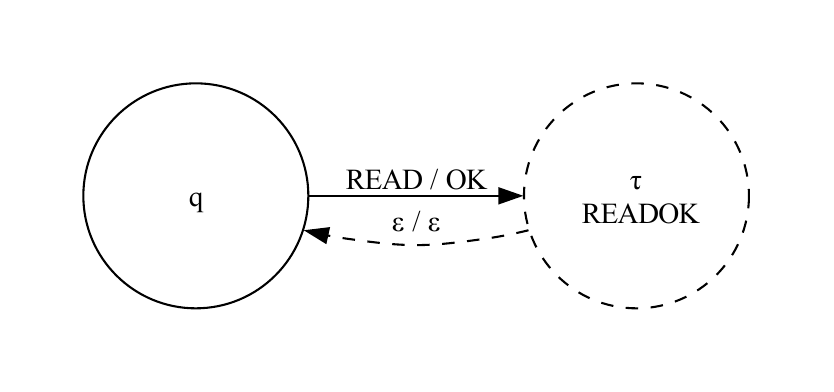}
\caption{Figure a) shows an authenticated and b) an unauthenticated state. Both figures show the state from the learned Mealy machine left and the same with an introduced \textit{implicit state proposition} to the right.}
\label{fig:TempProp}
\end{figure*}

\subsection{Conditions and Propositions}
%\noindent We define a condition as a connection between the propositions of our interest and the transitions in a Mealy machine. In each state, some propositions become true and some become false. We make the connection between the inputs and outputs on each incoming transition with gaining and losing a proposition (becoming true and false, respectively). Thereby, a condition is matched to each transition one-by-one (without a history) -- if a condition matches a transition, the target state gets the specified property.
\noindent Since we strive to automatically annotate Mealy machines with propositions, we need a rule set to define when a state gains or loses a proposition (i.e. it becomes true or false in that state). We therefore define conditions we can match with the Mealy machine's transitions. 
These conditions consist of one or more propositions we want to add or remove to a state and the input and output labels we compare the transitions with to evaluate a match.
If a condition matches we add or remove the addressed proposition to or from the respective transition's target state. We  evaluate each condition with each transition independently, there is no history or statefulness in the evaluation process. If multiple conditions match a transition, its target state gains or loses all propositions that are part of the condition (i.e., the result is a union of all matching conditions).

\begin{definition}[Condition]
\label{def:cond}
$\\$We define a condition
$c \in C$ as a triplet of proposition sets $\langle S_P,S_\sigma,S_\omega\rangle$. $S_P$ is a power set of $P$ ($S_p \in 2^P$), where $P$ denotes the set of all propositions assigned to a Mealy machine. $S_\sigma$ is an input power set ($S_\sigma \in 2^\Sigma$) and $S_\omega$ an output set ($S_\omega \in 2^\Omega$)  according to the semantics of the examined system protocol and properties of interest. 
\end{definition}
Since we are using power sets, a condition could determine multiple propositions ($p \in P$) to be set by one of multiple inputs ($\sigma \in \Sigma$) producing one of multiple outputs ($\omega \in \Omega$). We define a match function for a condition 
by looking at each transition. If the input label of a
transition 
matches one element of the input set of the condition, and analogously the output label 
of the same transition
matches one element of the output set,
then the match function returns true.

\begin{definition}[Condition Match Function]
\label{def:match}$\\$
The match function for condition $c$ at state $q$ returns true when the input $\sigma$ is part of the set $S_\sigma$ of condition $c$ and the output label 
$\omega=\lambda(q,\sigma)$ is part of the output set $S_{\omega_c}$ of the same condition. The function returns false otherwise.
$\\ 
\mathrm{match}_c(q,\sigma) := 
\begin{cases}
\top & \text{if } \big(\sigma \in {S_\sigma}_c\big)\land\big(\lambda(q,\sigma) \in {S_\omega}_c\big),\\
\bot & \text{otherwise}.
\end{cases}$

\end{definition}
The match function gives us the verdict if a transition (uniquely identified by its source state $q$ and its input $\sigma$) matches a condition $c$.
Using this match function, we can determine if a target state of such transition gains additional propositions (i.e. setting propositions from false to true) stated in a condition $c$. We define that, iff a transition (uniquely identified by $q$ and $\sigma$) matches the condition (i.e., the match function for $c$,$q$, and $\sigma$ returns true), we add the propositions in $S_P$ of $c$ the target state ($q\prime:=\delta(q)$). In the following definitions, we use $P(q)$ as formal notation for the set of propositions of a state $q$. Therefore, we define a set of proposition gain conditions as
\begin{definition}[Proposition Gain Conditions]$\\$
We define proposition gain conditions
$C_g$ as a set of conditions $c$ where the target state of a transition $q\prime = \delta(q)$ gets additional propositions $S_p$ defined in $c$ (relative to the origin $q$) if the condition matches the transition ($\mathrm{match}_c(q,\sigma)=\top$). These propositions are then false in $q$ but true in $q\prime$.
$\\C_g := \{c \in C \mid  \text{gain}_c\}$ and 
$\\ \text{gain}_c(q,\sigma):=
\begin{cases}
P(\delta(q)) \gets P(q) \cup S_p & \text{if } \mathrm{match}_c(q,\sigma),\\
P(\delta(q)) \gets P(q) & \text{otherwise.}
\end{cases}$
\end{definition}
Analogously, we define a set of loss conditions that determine when a target state $q\prime$ of a transition loses a set of propositions $S_p$ (propositions that become false).
\begin{definition}[Proposition Loss Conditions]$\\$
We define proposition loss conditions
$C_l$ as a set of conditions $c$ , where each $c$ specifies a set of propositions $S_p$ to be subtracted from a transition's target state $q\prime = \delta(q)$ relative to the source state $q$ if the condition matches the transition ($\mathrm{match}_c(q,\sigma)=\top$). These propositions are then true in $q$ but false in $q\prime$.
$\\C_{l} := \{c \in C \mid  \text{loss}_c\}$ and $\\\text{loss}_c(q,\sigma):=$
$\begin{cases}
P(\delta(q)) \gets P(q) \setminus S_p & \text{if } \mathrm{match}_c(q,\sigma),\\
P(\delta(q)) \gets P(q)  & \text{otherwise.}
\end{cases}$
\end{definition}

\subsection{Implicit State Propositions}
We may not be able to assign all the propositions in our CPM to a state in the Mealy machine using the above annotation algorithm (process), because we may have
to specify \textit{certain} properties that are related to labels on the transitions alone and not attributable to a state. 
%So, we have to create new states (which we call implicit states) and assign those propositions to these implicit states.
For an example, a state $q$ contains an operation (we want to attribute with proposition $p$) as a self-loop transition. Before performing the operation, $p$ should be false, afterwards it should become true. In such a case it is undecidable if state $q$ has proposition $p$ or not. It might, however, still be needed to check if $p$ does not occur in combination with some other proposition (e.g., a read must only occur if authentication is also set -- see Property P\ref{prop:auth} in Section \ref{sec:check:rules}). 
To resolve this, we define \textit{implicit} or \textit{implicit ($\tau$)} propositions. That is, propositions that are true only in a single, implicit state. 
To realize this, we split each transition (in the example above, the self-loop with $p$) into two parts: one transition leading from $q$ to a newly introduced implicit ($\tau$) state and one transition leading from the $\tau$ state to the original target state. We then label the transition towards it with the original input ($\sigma \in \Sigma$) and output ($\omega\in\Omega$) and the transition to the original transition's target state with an empty input and output ($\sigma\gets\varepsilon, \omega\gets\varepsilon$). These two transitions are the only ones connecting the $\tau$ state to the graph. This mechanism formally makes the resulting construct a Mealy-style subsequential transducer with partially defined transition and output functions \cite{rocheDeterministicPartofspeechTagging1995}. Every $\tau$ state has only a defined transition and output function for the emptiness symbol $\varepsilon$ as input, so not every state $q \in Q$ has a defined transition for every input $\sigma\in\Sigma$.
If the split transition matches an implicit condition ($c_\tau \in C_\tau$ - see Definition \ref{def:tempcond}), we also label the $\tau$ state with the condition's proposition set ($P(q_\tau)\gets c[S_p]$).
Figure \ref{fig:TempProp} gives an illustrative example of this mechanism: the two diagrams in \ref{fig:TempProp}a) show a state that possesses the $AUTH$ proposition, while the two in \ref{fig:TempProp}b) show one that does not. In both cases, we introduce an implicit state ($\tau$) to gain a checkable proposition for a successful read ($READOK$). It has this implicit state proposition and inherits all other propositions from $q$. This shows the purpose of such implicit state propositions. Given the property that one has to be authenticated to read a resource, the \ref{fig:TempProp}a) diagrams show a valid operation, while the \ref{fig:TempProp}b) violate that property inside the $\tau$ state. Using automata-based model checking, we cannot check for these cases without implicit state propositions, as we need a state where $READOK$ is set.

We therefore define a transition split function that creates a new implicit state, splits the original transition in two parts (from the origin to the implicit and from the implicit to the target state), and copies the implicit state propositions from the source state. We further create a set of implicit states $Q_\tau$ that do not contain the externally visible states of the classic Mealy machine part.
\begin{definition}[Transition Split Function]
\label{def:split}
The transition split function creates a new implicit state $q_\tau$, turns a transition $t (q\xrightarrow[\omega]{\sigma}q\prime)$ into two transitions 
$t\prime (q \xrightarrow[\omega]{\sigma} q_\tau)$ and $t\prime\prime(q_\tau \xrightarrow[\varepsilon]{\varepsilon} q\prime)$, and labels $q_\tau$ with the propositions from $q$:
    $\\f_\tau(q,\sigma,\omega,q\prime):= (Q\gets Q\cup\{q_\tau\},Q_\tau\gets Q_\tau\cup\{q_\tau\}, T\gets T\setminus \{(q,\sigma,\omega,q\prime)\}
    \cup\{ (q, \sigma, \omega, q_\tau), (q_\tau, \varepsilon, \varepsilon, q) \},P(q_\tau)\gets P(q)$.
\end{definition}
For the structure with split transitions, we then define a set of implicit state proposition gain conditions that will add a set of extra propositions $S_p$ to newly introduced $\tau$ states (we check this by using the implicit state set $Q_\tau$, we just introduced above) if there is a matching implicit condition $c_\tau$, similar to the standard proposition gain conditions.
\begin{definition}[Implicit State Proposition Gain Conditions]
\label{def:tempcond}$\\$
We define the set of implicit state proposition gain conditions $C_\tau$ as a set of conditions  $c$ where the target state of a transition $q\prime = \delta(q)$ gets additional properties $S_p$ defined in $c$ (relative to the origin $q$) if the condition matches the transition ($\mathrm{match}_c(q,\sigma)=\top$) and that target state is an implicit state ($q\prime \in Q_\tau$)
$\\C_\tau := \{c \in C \mid  \tau_c\}$ and $\\ \tau_c(q,\sigma):=
\begin{cases}
P(\delta(q))\gets P(\delta(q)) \cup S_p & \text{if } (\mathrm{match}_c(q,\sigma) \land \\ & (\delta(q) \in Q_\tau)),
\\P(\delta(q))\gets P(\delta(q)) & \text{otherwise.}
\end{cases}$
\end{definition}
That means that the same mechanism as in $C_g$ applies to $\tau$ states via $C_\tau$. Since the implicit state propositions should only be valid in the implicit state explicitly mentioned in $C_\tau$, we do not need loss conditions for implicit state propositions.

\subsection{CPM Composition}
Finally, we define a Context-based Proposition Map as a set of the three formerly defined sets.
\begin{definition}[Context-based Proposition Map]$\\$
A Context-based Proposition Map (CPM) is the set of the three sets for proposition gain conditions $C_g$, proposition loss conditions $C_l$, and implicit state proposition gain conditions $C_\tau$
$C:= \{C_{g},C_{l},C_\tau\}$
\end{definition}
We use these proposition maps (cf. Figure \ref{fig:Process}) to a) label Mealy machines with propositions (Section \ref{sec:check:AP}) and b) populate generic properties with protocol-specific propositions (Section \ref{sec:check:rules}).

\section{Annotating Mealy Machines with Propositions}
\label{sec:check:AP}

\begin{figure}[t]%[ht!]
	    \includegraphics[width=\linewidth]{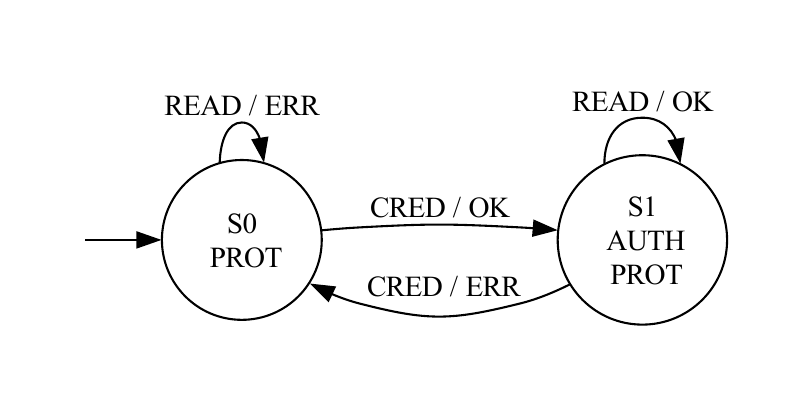}
\caption{A graphical AMM representation of an example for a simple authenticated file read mechanism.}
\label{fig:ExampleAMM}
\end{figure}

Since the Mealy machines used in our approach lack propositions that are needed for (Kripke-structure-based) model checking, we use a CPM to annotate them. 
To this end, 
we first apply the conditions in set $C_g$ to all transition labels of the 
model. The \textit{target} state of each transition gains $S_P$ if the transition match $c$ according to matching function (Definition \ref{def:match}). We further assume that propositions will propagate over transitions, unless there is a condition that prevents this. For instance, a system that is in an authenticated state will stay so, even if the state changes, until there is an event that triggers a de-authentication. These preventive conditions are the ones we denote in the proposition loss condition set $C_l$. We therefore, for each transition $t \in T$, add the propositions of the source state to the ones of the target state, except if the transition matches an element of the set of lose conditions: ($\forall t \in T \mid P(q\prime_t):=P(q\prime_t)\cup P(q_t) \text{ if } \neg \text{ loss}_c(t[q],t[\sigma])$). Since now more states have a certain proposition set, this proposition may propagate even further. We therefore re-iterate this process with the updated model until no state gains a new proposition, which means the propagation is complete. Algorithm \ref{alg:m-to-mk} shows a pseudocode of the labeling algorithm to annotate Mealy machines with propositions.
\begin{algorithm}
\caption{Mealy Machine Annotation}
\label{alg:m-to-mk}
\begin{algorithmic}[1]
\Statex \textrm{//Step 1: Parse Mealy Machine}
\ForAll{lines $\ell$ in DOT file of the Mealy machine}
	\If{$\ell$ denotes state}
        \State $Q \gets Q \cup (q \text{ from } \ell)$
        \If{$\ell$ declares initial state} 
            \State $q_0 \gets$ \textsf{($q$ from $\ell$)} 
        \EndIf
    \EndIf
    \If{$\ell$ denotes transition}
        \State $t \gets$ (\textsf{$(q,q\prime,\sigma,\omega)$ from $\ell$}); $T \gets T \cup t$
    \EndIf
\EndFor
\Statex \textrm{//Step 2: Distribute Initial Propositions using the Gain Conditions}
\ForAll{transitions $t$ in $T$}
    \ForAll{conditions $c_g$ in $C_g$}
        \If{match($c_g, t$)}
            \State $(q,i,o,q\prime) \gets t, P(q\prime)\gets P(q\prime)\cup c[S_p]$ 
        \EndIf    
    \EndFor
\EndFor
\Statex \textrm{//Step 3: Iteratively Propagate Propositions unless there is a Loss Condition}
\Repeat
  \State changed $\gets \bot$
  \ForAll{{transitions $t$ in $T$}}
      \ForAll{conditions $c_l$ in $C_l$}
        \ForAll{properties $p$ in $P(t[q])$}
            \If{$(\neg (p \subseteq P(q\prime)) \land (\neg match(c_l,t))$}
                \State $P(q')\gets P(q')\cup P(q)$
                \State changed $\gets \top$
            \EndIf
        \EndFor
    \EndFor
  \EndFor
\Until{\textbf{not} changed} 
\end{algorithmic}
\end{algorithm}
Eventually, we have an automaton annotated with all propositions according to the defined conditions. 
As a result, we have labeled states like in a Kripke structure. Since we still maintain the Mealy machine labels of transitions, this resulting \textit{annotated Mealy machine (AMM)} can be seen as combined Mealy-Kripke structure, effectively extending the classical Mealy machine definition with a set of propositions $P$ and a labeling function $\mathcal{L}$.
\begin{definition}[Annotated Mealy Machine]$\\$
\label{def:AnnotMealy}
We define an annotated Mealy machine (AMM) as
$\mathcal{MK}:= (Q,\allowbreak q_0,\allowbreak \Sigma, \allowbreak \Omega, \allowbreak,\delta,\allowbreak \lambda,\allowbreak P, \allowbreak \mathcal{L})$, with $Q$ being the set of states, $q_0$ the initial state, $\Sigma$ the input alphabet, $\Omega$ the output alphabet, $\delta$ the transition function, and $\lambda$ the output function of the Mealy machine. Finally, $P$ denotes the set of propositions and $\mathcal{L}$ the labeling function that attributes propositions $p\in P$ to the states. 
\end{definition}
Figure \ref{fig:ExampleAMM} shows an example AMM that has been generated using the example Mealy machine (Section \ref{sec:check:AP:CPM:Example}) and CPM. We thereby assume that the resource in the Mealy machine is protected, hence all states get the $PROT$ proposition. We 
%%%%%%%%%%%%%%%%%%%%%
%\TODO{Refer to Galois lattices~\cite{edwards_galois_1984,galois_sur_1846}.} 
%%%%%%%%%%%%%%%%%%%%%
The AMM's proposition set is implicitly defined as the set of all propositions that occur in the \textit{gain} conditions $C_g$ of the proposition map of a corresponding CPM. For building the set we can ignore $C_l$, because if it contains an additional proposition, it is only subtracted and will therefore not show up in the resulting AMM. We can also ignore $C_\tau$, as it only contains the implicit state propositions that are also not visible.
\begin{definition}[Annotated Mealy Proposition Set]$\\$
We define the proposition set $P$ of an annotated Mealy machine as the set of all $p$ that occur in one of the gain conditions of a corresponding CPM.
$P:=\{p \mid\exists(S_p,S_\sigma,S_\omega)\in C_g; p \in S_p\}$
\end{definition}

$\mathcal{L}$ uses the corresponding CPM's functions. We explicitly emphasize that $\mathcal{L}$ resembles the labeling function of a Kripke structure (as it labels states) not an LTS' labeling function that in contrast labels transitions. The latter part is handled by the output function $\lambda$ like in a standard Mealy machine.
Therefore, CPMs enable to check different complex protocols with the same generic properties, as they instantiate the properties by populating it with the necessary (but protocol-specific) propositions. Formally, that means that we execute the \textit{gain, loss} and $\tau$ functions of a proposition map $C$. 
\begin{definition}[Annotated Mealy Labeling Function]$\\$
We define the labeling function $\mathcal{L}$ of an annotated Mealy machine as the sequential execution of the $gain$, $loss$, and $\tau$ functions of an associated CPM $C$.
$\mathcal{L}(C,q,\sigma):= \allowbreak{gain}_c(q,\sigma)\allowbreak \oplus {loss}_c(q,\sigma)\allowbreak \oplus \tau_c(q,\sigma)$
\end{definition}

\begin{figure}[t]%[ht!]
	    \includegraphics[width=\linewidth]{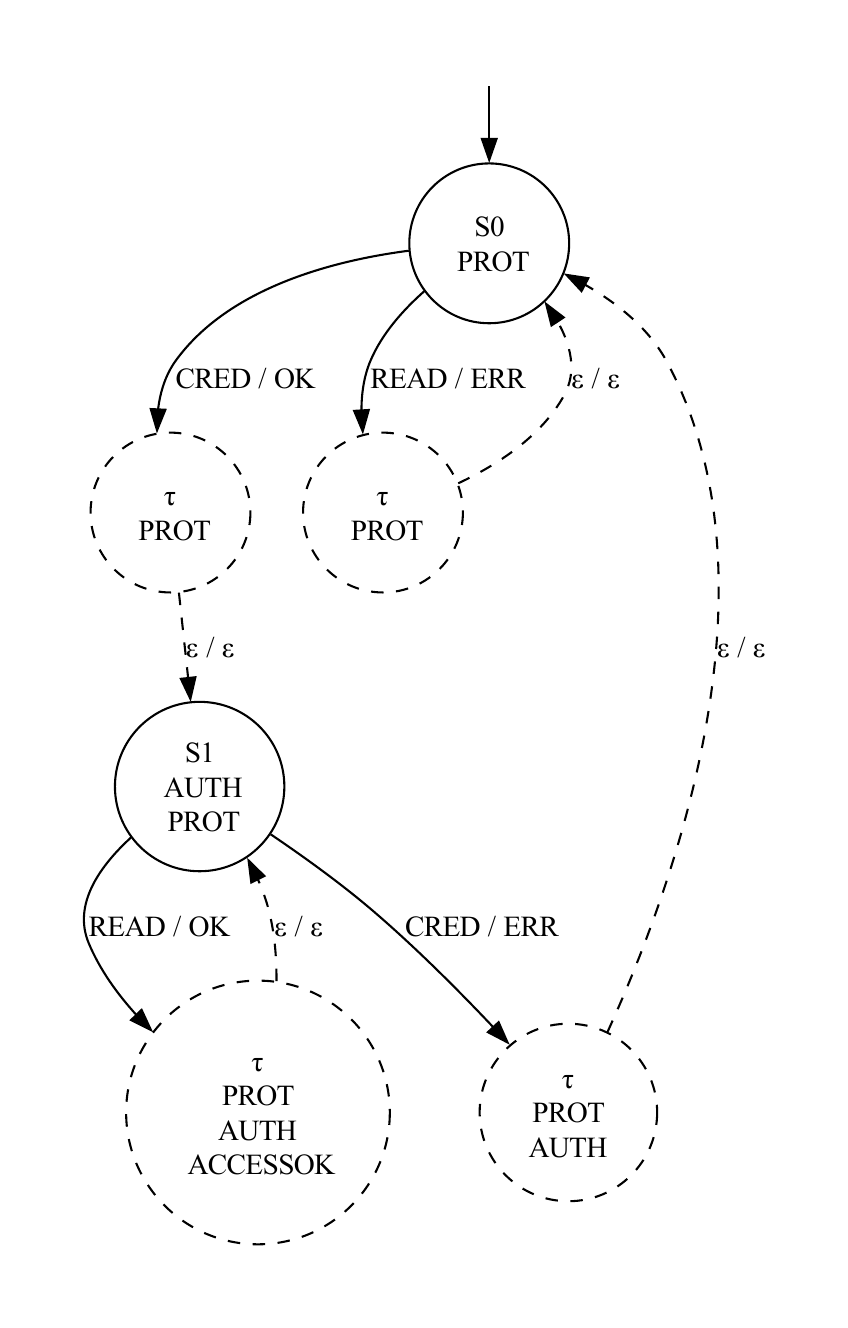}
\caption{A graphical AMM representation of an example for a simple authenticated file read mechanism with visible implicit states.}
\label{fig:ExampleAMMtau}
\end{figure}

\subsection{Annotating Implicit State Propositions}
\label{sec:gen:temp}
We only use the implicit state propositions from $C_\tau$ in the translation to Rebeca code (Section \ref{sec:gen}), they are not directly visible in the AMM.
For any input and output matching $c \in C_\tau$, we implement the split translation function (Definition \ref{def:split}) by splitting the respective transition (labeled \textit{IN/OUT}) into two parts: an implicit transition (labeled \textit{IN/OUT}) from the source state ($q$) to a new implicit ($\tau$) state and an empty-input transition (labeled $\varepsilon$/$\varepsilon$) from $\tau$ the target state ($q\prime$, which may or may not be equal to $q$). The generated $\tau$ state does not have any transitions than these two fragments of the original one.
 We then let state $\tau$ inherit all propositions from $q$ and additionally set the proposition $p$ that is defined in $c$, which is then (only) true in $\tau$. Since we do not alter the target state, there is automatically no further inheritance. 
To demonstrate this on our illustrative example, every transition gets an implicit ($\tau$) state, that gets all propositions from the source state. Additionally, the self-transition of $S1$ labeled $READ/OK$ gets the $ACCESSOK$ proposition, since this condition is stated in the example CPM. Figure \ref{fig:ExampleAMMtau} shows a representation of that machine with visible implicit states. 

\section{Generating Rebeca Code}
\label{sec:gen}
After annotating the Mealy machine
with propositions, a fully automated procedure implemented in Java translates the resulting 
AMM to Rebeca code, where it can be subjected to model checking. This section (formally) describes a Rebeca template and the transformation process.
 
\subsection{Definition of a Rebeca Template}
\label{sec:check:template}
In the AMM we see input and outputs of each transition. In the corresponding Rebeca code we have two actors:
an \textit{environment} (an external entity that provides inputs and collects outputs) and a \textit{system} (the entity whose behavior is defined by the AMM). 
\begin{definition}[Rebeca Template]$\\$
\label{def:template}
We define a Rebeca template $\mathcal{R}$ as the parallel execution of a \textit{system} actor
$r_{sys}$ that obtains the annotated Mealy machine's behavior and an \textit{environment} actor $r_{env}$ that serves as an input giver.
$\mathcal{R}:=r_{sys}||r_{env}$ 
\end{definition}

The system actor $r_{sys}$ holds a state variable with an identifier of the current state and
one variable for each proposition in the AMM's set of propositions, including implicit state propositions (e.g., $auth, accessok$) .The only known actor is the \textit{environment} and the method identifiers are equal the set of input symbols (e.g., $read(),cred()$) plus an additional \textit{request} method that serves as a calling and reset point.

\begin{definition}[System Actor]$\\$
\label{def:sysact}
We define the \textit{System Actor} $r_{sys}$ as an actor with 
 a state variable as an identifier of the current state ($q\in Q$) from the annotated Mealy machine, initially set to $q_0$), each one boolean state variable for the annotated Mealy machine's set of propositions $P$, and each one boolean state variable for the implicit state propositions in the context-based proposition map. Its known actor is the environment actor $r_{env}$. It possesses a method $f_\sigma$ for each input Symbol of the annotated Mealy machine $\sigma \in \Sigma$ and a \textit{request} method that handles the calls of the system and resets the implicit state propositions.
$r_{sys}:=\langle V_{sys},M_{sys},K_{sys}\rangle$, with the system variables $V_{sys}:=\{q\}\cup P \cup \{p \mid\exists(S_p,S_\sigma,S_\omega)\in C_\tau: p \in S_p\}$, the system methods $M_{sys}:=\{f_\sigma \mid \sigma \in \Sigma\}\cup\{req()\}$, and the environment as known actor $K_{sys}:=\{r_{env}\}$.
\end{definition}

The environment actor contains the 
\textit{system} (Definition \ref{def:sysact}) as the only other known actor. The set of method identifiers consists of the output alphabet (e.g., $ok(),err()$).

\begin{definition}[Environment Actor]$\\$
\label{def:envact}
We define the environment actor $r_{env}$ as an actor with the \textit{system} as known actor ($r_{sys}$). The set of method identifiers consists the output alphabet ($f_\omega \mid \omega \in \Omega$).
$r_{env}:=\langle V_{env},M_{env},$ $K_{env}\rangle$), with the implicit state propositions as variable set 
$V_{env} :=\{\}$
, the environment methods $M_{env}:=\{f_\omega \mid \omega \in \Omega\}$ and the system as known actor ($K_{env}:=\{r_{sys}\}$).
\end{definition}

\subsection{Template Semantics}
The main part of the Mealy machine's logic is modeled in the \textit{system} actor, while the \textit{environment} actor can be seen as an external input giver to the system.
To start a Rebeca model built with the defined template, \textit{environment's} constructor starts the \textit{request} message server on the \textit{system}. There, we reset all state variables (which represent the implicit state propositions from $C_\tau$) to false and non-deterministically
send messages corresponding to each possible input symbol
by calling a method $f_\sigma$ from $M_{sys}$ via sending the message $\langle r_{sys},r_{sys}, f_\sigma\rangle$, with a randomly picked $\sigma \in \Sigma$. Therefore, a local variable \textit{data} gets a random value from 0 to $|M_{sys}|=|\Sigma|$ and a switch/case calls one distinct methods from $M_{sys}$ (e.g., $cred()$).
In each of the methods of $M_{sys}$ ($f_\sigma \in M_{sys}$) four operations are performed: a) we set the current state identifier ($q \in V_{sys}$) according to the AMM's transition function $\delta(q,\sigma)$ (e.g., from $0$ to $1$),
b) we set  all property variables ($p \in V_{sys}$) according to the labeling function ($\mathcal{L}$) in the AMM for the target state\footnote{Note: the Method $f_\sigma \in M_{sys}$ has a direct, bijective representation in the set of Input Symbols ($\sigma \in \Sigma$): $M_{sys} \leftrightarrow \Sigma$. Additionally having the state variable, we can therefore directly use the transition ($\delta$) and output ($\lambda$) functions of the AMM.} (e.g., setting $auth$ to $\top$),
c) we set a state variable corresponding to a \textit{implicit state proposition} to true, if there is a respective entry in the CPM ($C_\tau$ - e.g., setting $accessok$ to $\top$, if $read()$ occurs in state $S1$).
d) we call a message server of the \textit{environment} actor from $M_{env}$ ($f_\omega \in M_{env}$) that corresponds to the respective output in the respective current state determined by AMM's output function $\omega=\lambda(q,\sigma)$ (e.g., calling $r_{env}$'s $ok()$ method).
This means sending the message $\langle r_{sys},r_{env},f_\omega\rangle$. 

Each of these output message servers $f_\omega \in M_{env}$ (mapping to a $\omega \in \Omega$ from the AMM)
calls the \textit{request} method (with the message $\langle r_{env},r_{sys},{req}\rangle$) to start another interaction cycle. This includes, again to reset all implicit state propositions (i.e., resetting the respective state variables in \textit{system} to $\bot$). This assures that each implicit state proposition is present in a state for the model checker, but not maintained further, as it is not part of any state of the AMM.

\subsection{Rebeca Code Transformation}
To transform the AMM into Rebeca code according to the template, we use an automated translation process (which we implemented in Java) that involves basically four steps before writing the actual code. We describe these steps informally before presenting the whole algorithm in formal (pseudocode) notation and giving the results from our running example in the last paragraph. Step 1 is parses the DOT file of the machine line by line and create processable data structure for the containing states and transitions, as well as identifying the initial state. Step 2 determines for both actors the variables ($V_{sys},V_{env}$) and known counterparts ($K_{sys},K_{env}$). Since the template involves two communicating actors, the counterparts are trivial, as they have to simply know each other. 
 The state variables of the system actor ($V_{sys}$) are determined by the propositions (one boolean for each $p \in P$), plus an integer to keep track of the system state. The environment's variables ($V_{env})$) consist of all implicit state propositions (every $p$ that occurs in one of the propositions sets in the implicit conditions ($P_\tau \in 2^P$). Step 3 creates the constructor and the request methods of the environment actor. The request method ($req$) is the driving function and randomly calls an input method from the system actor ($f_{\sigma}$) and resets all implicit state proposition variables ($p\in V_{env}$) to false, while the constructor simply calls request to start the model. Step 4 parses the transition structure (from Step 1) adds a  method for each input and output encountered: an input method $f_\sigma$ into the system actor $V_{sys}$ for every input $\sigma \in \Sigma$ and an output method $f_\omega$ into the environment actor $V_{env}$ for every output $\omega \in \Omega$. 
Each input method contains a conditional on the current state $q$ and depending on that then a) sets the state variable to $q\prime$ of the transition, b) checks for a proposition delta between $q$ and $q\prime$ and modifies all $p \in V_{sys}$ accordingly and c) calls the output method $f_\omega$ that matches $\omega$ in the transition.
Each output method contains a call to $req$ at the end.
This implements the implicit transition split described in Section \ref{sec:check:AP:CPM}: 
The calls of $f_\sigma$ and $f_\omega$ represent the first part (transition $q \xrightarrow[\omega]{\sigma} q_\tau$, setting the implicit state propositions), while internal call of $req$ from each $f_\omega$ represents the second part (transition $q_\tau \xrightarrow[\varepsilon]{\varepsilon} q\prime$, resetting the implicit state propositions). Further, if a condition for an implicit state proposition matches the output that needs to be distinguished by the input, a parameter for the caller ID (ID$_\sigma$) is added along with a switch for that parameter, that sets the variable for the implicit state proposition ($(p_\tau \in V_{env}) \gets \top$). This ID is the same the $req$ methods uses to call the respective $f_\sigma \in M_{sys}$. The ID parameter and switch is omitted if there is less than two options for a calling $f_\sigma$ or the lack of matching implicit state propositions.

More formally, Algorithm \ref{alg:mk-to-rebeca} lists a pseudocode that transforms an AMM ($\mathcal{MK}$, see Definition \ref{def:AnnotMealy}) into Rebeca code according to the given template ($\mathcal{R}$), see Definition \ref{def:template}.
\begin{algorithm}
\caption{Annotated Mealy Machine to Rebeca Code Transformation}
\label{alg:mk-to-rebeca}
\begin{algorithmic}[1]

\Statex \textrm{//Step 1: Parse AMM}
\ForAll{lines $\ell$ in DOT file of the AMM}
	\If{$\ell$ denotes state}     
        \State $Q \gets Q \cup (q\text{ from }\ell), P(q)\gets (P \text{ from }\ell)$
        \If{$\ell$ declares initial state} 
            \State $q_0 \gets$ \textsf{($q$ from $\ell$)} 
        \EndIf
    \EndIf
    \If{$\ell$ denotes transition}
        \State $t \gets$ (\textsf{$(q,q\prime,\sigma,\omega)$ from $\ell$}); $T \gets T \cup t$
    \EndIf
\EndFor
\Statex \textrm{//Step 2: Create Variables and Known Actors}
\State $V_{sys} \gets \{\, (p,\bot) \mid p \in P\} \cup {(q\_id,0)} \cup
\{\, (p_\tau,\bot) \mid \exists\ i,o:\ \langle i,o,p\rangle \in C_\tau\}$ \Comment{$q\_id$:tracker for q, $p \text{ from }c_\tau$: implicit state propositions}
\State $V_{env} \gets \{\}$
\State $K_{sys} \gets $ ENV
\State $K_{env} \gets $ SYS
\Statex \textrm{//Step 3: Create environment constructor and system actor's request method}
\Function{\textbf{Constructor}}{}\Comment{in $M_{env}$}
    \State \Call{SYS.REQ}{\text{}}
\EndFunction
\Function{REQ}{\text{}}\Comment{in $M_{sys}$}
    \State $\forall p_\tau \in V_{sys}: p_\tau \gets \bot$
    \State data $\gets (? \in \{0..(\lvert\Sigma\rvert-1)\})$
    \Switch{data}
        \Case{0} \Call{$f_{\sigma_0}$}{\text{}} $\in M_{sys}$
        \Case{\text{...}}   ...        
        \Case{$\lvert\Sigma\rvert-1$} \Call{$f_{\sigma_{\lvert\Sigma\rvert-1}}$}{\text{}} $\in M_{sys}$
    \EndSwitch
\EndFunction
\State $M_{sys} \gets \{\text{REQ()}\}$
\State $M_{env} \gets \{\text{Constructor()}\}$
\Statex \textrm{//Step 4: Parse Transitions and Create Methods}
\ForAll{$t \in T$}
    \Function{$f_\omega$}{\text{}} \Comment{Output functions in $M_{env}$}    
        \State \Call{req}{\text{}}
    \EndFunction
    \State $M_{env} \gets M_{env} \cup f_\omega(\text{})$ 
    \Function{$f_\sigma$}{\text{}} \Comment{Input functions in $M_{sys}$}
        \ForAll{$s \in Q$}
            \If{$s=t[q]$}
                \State $V_{sys}[q] \gets t[q\prime]$
                \ForAll{$p \in P$}
                    \If{$q[p] \neq q\prime[p]$}
                        \State $p \gets q\prime[p]$
                    \EndIf
                \EndFor
                \If{match($c_\tau \in C_\tau$, t)}
                    \State $V_{sys}[p_\tau]\gets \top$
                \EndIf        
                \State\Call{$f_{\omega_t}$}{\text{}}
            \EndIf
        \EndFor
    \EndFunction
    \State $M_{sys} \gets M_{sys} \cup f_{\sigma}$
\EndFor
\end{algorithmic}
\end{algorithm}

\begin{figure}[t]%[ht!]
        \includegraphics[width=\linewidth]{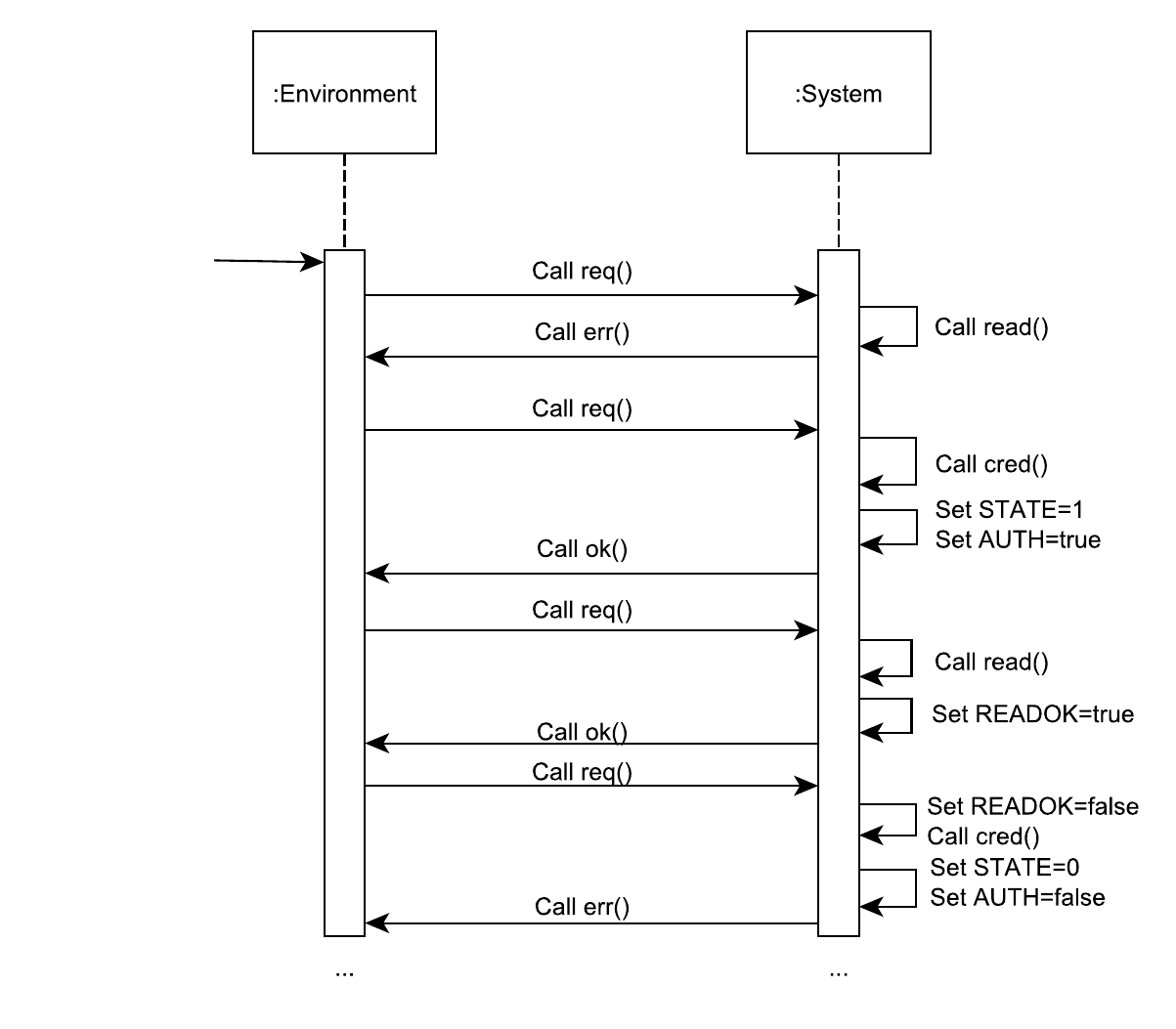}

\caption{An UML sequence diagram of its Rebeca representation (right), as outlined in Section \ref{sec:gen}.
}
\label{fig:MealyRebeca}
\end{figure}

Going through the described process with our example from Section \ref{sec:check:AP:CPM:Example}, we receive the following Rebeca model $\mathcal{R}$, as defined in Section \ref{sec:check:template}:
 $\\r_{sys}:=\langle$$V_{sys}:=\{state,prot,auth,accessok\},$ $M_{sys}:=\{req(),read(), cred()\allowbreak\}$,
 $K_{sys}:=\{r_{env}\}\rangle$.
 $\\r_{env}:=\langle$$V_{env}:=\{\},$ $M_{env}:=\{ok(),\allowbreak err()\},$$K_{env}:=\{r_{sys}\}\rangle$.
Figure \ref{fig:MealyRebeca} shows a graphical representation of this example, while Listing \ref{lst:RebEx} shows the resulting Rebeca code. Please note that in the Rebeca code we added an additional start function, as the constructor can only call own (\textit{self}) functions and that the system actor has an error (\textit{ERR}) message server that is just used for bug fixing models.

\begin{figure}[htp!]
\centering
\begin{lstlisting}[label=lst:RebEx,caption=Rebeca code from the illustrative example from Section \ref{sec:gen}.,language=rebeca,escapechar=|]
reactiveclass ENVIRONMENT(3) {
knownrebecs {
	SYSTEM system;
}
ENVIRONMENT() { self.start(); }
void start() { system.req(); }
msgsrv err(){ system.req(); }
msgsrv ok(){ system.req(); }
}

reactiveclass SYSTEM(3) {
knownrebecs {
	ENVIRONMENT environment;
}
statevars {
	boolean prot, auth;
	int state;
	boolean error, accessok;
}
SYSTEM() {
	prot=false;
	auth=false;
	state=0;
}
msgsrv req() {
	error=false;
	accessok=false;
	int data =?(0,1);
	switch(data) {
		case 0: self.cred(); break;
		case 1: self.read(); break;
		default: self.ERR();
	}
}
msgsrv ERR(){
	error = true;
}
msgsrv read(){
	if(state==0) {
		state=0; 
		environment.err();
	} else
	if(state==1) {
		state=1; 
		accessok=true;
		environment.ok();
	}
}
msgsrv cred(){
	if(state==0) {
		auth=true; 
		state=1; 
		environment.ok();
	} else
	if(state==1) {
		auth=false; 
		state=0; 
		environment.err();
	}
}
}
main {
	ENVIRONMENT environment(system):();
	SYSTEM system(environment):();
}

\end{lstlisting}
\end{figure}

\subsection{Altering the Model} 
\label{sec:check:alter}
Since in Rebeca a model's behavior is defined similar to a programming language, altering it is trivial. Furthermore, Rebeca possess an operator for non-determinism (the $?$-operator). On the one hand, we can manually add any arbitrary fault that might be interesting to investigate and examine its impact on keeping the properties. On the other hand, this allows for re-introducing some non-deterministic behavior that is abstracted away during the learning process~\cite{sirjani_actors_2025}. For instance, we reintroduced timeouts that have been abstracted, by altering the template mentioned in Section \ref{sec:check:template}. We just introduce a \textit{timeout} server on the sender side and alter each input message server on the receiver to trigger a timeout with a certain probability, realized with a random ($?$) pick of an integer. If no timeout occurs, the logic continues. Otherwise, we reset the state to initial and call the timeout function on the sender, which sets an implicit state proposition for \textit{timeout}, to allow for checking for it. Since we can set the timeout probability to an arbitrary value, we can play with different scenarios to be checked. We can also convert the resulting Rebeca code back to a Mealy machine, so we possess a model that can contain timeouts and errors, just as the originally learned model would have. If the resulting Rebeca code contains non-determinism, the resulting model is not technically a Mealy machine, but a \textit{Mealy-styled non-deterministic state model}, for it defines multiple transitions for a given input per state to represent the different possibilities. Formally, we alter the Rebeca template by extending $M_{env}$ with a \textit{timeout} method, that sets a state variable ${timeout} \in V_{env}$ (to perform checks for timeouts if needed) and calls the \textit{request} method. Further, we alter each method $m_{sys} \in M_{sys}$ to randomly (with a settable parameter determining the probability) reset the state identifier $q$ to the initial state $q_1$ and call the \textit{timeout} method ($\langle r_{sys},r_{env},{timeout}\rangle$), instead of the output method ($M_{env}(\omega)$ derived from $\lambda(q,\sigma)$).

\subsection{Verifying the Code}
\label{sec:check:verify}
In order to assure the Rebeca code is faithful to the learned model, we check their equivalence. 
We generate the full state space of  the Rebeca model. For that, 
we run the Rebeca model checker (RMC) without any properties to check, so RMC does not stop at a violation. This implicitly means that RMC just checks for standard assertions (e.g., deadlocks, starvation, queue overflows), which should not occur by construction through the used Rebeca template. If it does not find a violation, RMC stops when it has explored the full state space \cite{khamespanahAfraEclipseBasedTool2023}.
This way the full state space is generated, which we automatically export to a Linear Transition System (LTS) representation in the GraphViz format. Based on a previous work in~\cite{marksteinerBlackboxProtocolTesting2025}, we then use an algorithm that collapses the state space LTS into a Mealy machine and check the bisimilarity of the converted Rebeca Mealy machine with the originally learned model. This an additional optional step to provide assurance that no errors have been introduced in the transformation process.

\section{Checking the Model}
\label{sec:check}
The Rebeca model checking tool (RMC) supports Linear Temporal Logic (LTL) to formulate properties. 
For security checking, these LTL formulas should describe security properties. Our approach allows us to use generic security properties, which we can concretize for specific automata using CPMs. A property consists of propositions, which the CPM defines (using semantics from the CPM's protocol context). This results in an AMM that includes the propositions, which the security properties we want to check consist of. In this section, we begin with defining four basic, well-established security requirements (authentication, confidentiality, privilege levels, and key validity) and for each  define a corresponding security property that we can check. Lastly, we briefly describe how to concretize the properties' propositions using CPMs in two use cases (eMRTDs and UDS). Section \ref{sec:eval} subsequently describes the practical application and evaluation in these two use cases.

\subsection{Generic Security Requirements}
\label{sec:reqs}
In this section, we take some well-established security properties~\cite{saltzer_protection_1975} that are also aligned with our case studies' standards documents (ICAO~\cite{international_civil_aviation_organization_machine_2021} and UDS~\cite{international_organization_for_standardization_road_2020}) and formulate them as checkable requirements in \textit{given-when-then}\footnote{Using \textit{given, when,} and \textit{then} as keywords in italics.} format~\cite{binamunguBehaviourDrivenDevelopment2023} \footnote{We use the terms MUST, MUST NOT, SHOULD, SHOULD NOT and MAY as defined by the Internet Engineering Task Force (IETF)~\cite{%brandner_key_1997,
leiba_ambiguity_2017}, using the same CAPITALIZED style.}. These requirements use attributes (informal traits of a requirement's subject), we later (in Section \ref{sec:check:rules}) formalize as propositions\footnote{Using \textsc{small capitals} style.} for defining logic properties. This definition was performed manually, however, given the generality of the properties and their automated instantiation via CPMs we demonstrate below, the process can be considered a one-time effort. If other (either specific or generic) security properties are needed, we recommend to first state a semi-formalized requirement (e.g., \textit{in given-when-then} format) analyze the components for attributable propositions and constitute the property to meet the requirement's semantics. Subsequently, the properties can be instantiated by populating the propositions with a concrete system context with a CPM.
%\footnote{To not collide with the \textit{Given}/\textit{when}/\textit{then} format in readability, we use \textit{italics} instead of capitals when using IETF-defined key words.}

\subsubsection{Authentication}
\label{sec:reqs:auth}
To establish the authenticity as a property, we must require \textsc{authentication} of a user before accessing \textsc{protected} resources (not every resource must be protected, some may be openly accessible -- this depends on the system's security design).
As a requirement, we define it as follows:
\Requirement{
\label{req:auth}
%Given the system is in a state with the proposition \textsc{authenticated} set to false, when a read operation on a \textsc{protected} resource occurs, then the operation must not return a positive response (\textit{read ok}).
\textit{Given} the system is not in an \textsc{authenticated} state, \textit{when} an \textsc{access} operation on a \textsc{protected} resource occurs, \textit{then} the operation MUST NOT return a positive response.
}
\subsubsection{Confidentiality}
\label{sec:reqs:conf}
To achieve confidentiality (also called secrecy in cryptographic contexts), most protocols allow for encryption (and may require it under certain conditions). We therefore recommend encryption for protected resource access:
\Requirement{
\label{req:conf}
\textit{Given} the system is in an arbitrary state, \textit{when} an \textsc{unsecured} \textsc{access} operation on a \textsc{protected} resource occurs, \textit{then} it SHOULD NOT  return a positive response.
}
\textsc{Unsecured} \textsc{access} thereby denotes a file operation without cryptographic security controls (i.e., encryption).

\subsubsection{Privilege Levels}
\label{sec:reqs:priv}
It is common practice for more complex systems and protocols to define more than one privilege level according to different roles and corresponding rights of an actor in order to provide access to more critical resources. This also applies for UDS, where this is implemented by different session and security levels~\cite{internationalorganizationforstandardizationRoadVehiclesUnified2020}. Likewise, for eMRTDs, sensitive data have to be additionally protected via terminal authentication.
\Requirement{
\label{req:priv}
\textit{Given} the level of \textsc{privilege} is not sufficient,  \textit{when} an \textsc{access} operation on a \textsc{critical} resource occurs, \textit{then} the operation MUST NOT be successful. A sufficient level of privileges also implies authentication.
}

\subsubsection{Key Validity}
\label{sec:reqs:key}
Another important property of secure connections is key validity and recency (i.e., no newer key must exist). Since we do not consider a time component in the automata, we do not use recency in the sense of freshness. 
An \textit{invalid key} in this sense is either a \textit{wrong key} or an \textit{old key} (not recent). 
\Requirement{
\label{req:key}
%\textit{Given} an \textsc{authentication} OR \textsc{secure access} operation, \textit{when} an actor is asked for a key, \textit{then} providing an \textit{invalid key} MUST NOT be successful.
\textit{Given} any operation that requires a key, \textit{when} an \textsc{invalid key} is provided, \textit{then} the key MUST NOT be accepted.
}

\subsection{Defining Generic Properties}
\label{sec:check:rules}
Since the requirements in Section \ref{sec:reqs} are generic, we can also define generic properties to check them. Generally, we use the \textit{given} and \textit{when} parts as a premise and the \textit{then} part as a consequence (except for the mapping from Requirement R\ref{req:key} to property P\ref{prop:key}, for which we trivially use \textit{no invalid keys} as a property).
These become more specific in conjunction with the CPMs, which define the meaning of propositions used in the property in the context of a specific protocol. If a proposition used in a property is not defined in the respective CPM, we have two options, which we both implement to leave the user the choice of a safer or a more convenient CPM definition: a) leave it undefined, which causes an error in the model checking and makes the user aware of this fact or b) define it as \textit{false}($\bot$) for the respective property. Option a) works as a \textit{strict} mode that prevents the user defining the CPM introduce errors by leaving propositions undefined. Option b) means that, for semantically sensible settings, our properties hold with undefined propositions, effectively ignoring the concerned proposition, so CPMs can be defined more quickly and easily without breaking the functionality. 
For the sake of readability, we restrict our definitions of \textsc{access} to read operations;  write, update, and similar operations can be defined the same way.

\subsubsection{Authentication}
\label{sec:check:rules:auth}
For Requirement R\ref{req:auth}, we define an authenticated state as a state with the proposition \textsc{authenticated} set to true. If this is not the case, 
when an access operation on a \textsc{protected} resource occurs, then the operation must not return a positive response.

\Property{
\label{prop:auth}
$\square ( \neg {AUTH}  \land  {PROT}  \rightarrow   \neg  {ACCESSOK})$\footnote{Afra does not support implications in the property file, therefore we use  $\square ( \neg ( \neg {AUTH}  \land  {PROT})  \lor   \neg  {ACCESSOK})$, instead.}.
}
 Here, we define ${AUTH}$ for \textsc{authenticated},  ${PROT}$ for \textsc{protected}, and ${ACCESSOK}$ for a positive response (i.e., a successful read operation). If we have a protocol with no protected resources, we can set $PROT:=\bot$, which means that reading is always allowed according to Property~P\ref{prop:auth}. 

\subsubsection{Confidentiality}
\label{sec:check:rules:conf}
For checking the confidentiality (Requirement R\ref{req:conf}), we assume the presence of unsecured and secured operations. For the \textit{access} operation as used in Section \ref{sec:check:rules:auth}, we therefore define two exemplary subsets: an \textsc{unsecured} read $UREAD$ and a \textsc{secured} read $SREAD$. 
Since $READ$ is a superset, in the respective context maps we then define that $READ$ will always be set along with $UREAD$ \textit{or} $SREAD$ is set (using the same conditions, i.e., $c_{READ} = c_{UREAD} \oplus c_{SREAD}$). 
Given requirement R\ref{req:conf}, we define that an $UREAD$ may not succeed when $PROT$ is true. Please note that we fail to apply this property to the UDS case, since there is no encrypted communication on the CAN bus that servers as a medium for the UDS protocol. While there are some proprietary and research solutions for encrypted CAN, none of them is widely adopted (i.e., used by a larger number of major OEMs). An industry standard gaining momentum, the AUTOSAR \textit{Secure On-board Communication (SecOC)}~\cite{autosar_specification_2024}, only provides authentication and integrity, but not confidentiality.
\Property{
\label{prop:conf}
$\square ({PROT} \rightarrow \neg \text{ }{UREADOK})$
}

\subsubsection{Privilege Levels}
\label{sec:check:rules:priv}
%We assume that there are areas that go beyond a basic protection level and that there additional protection measures (e.g., different type or level of authentication) have to be taken (Requirement R\ref{req:priv}). We define this higher protection level as \textit{critical} resources ($CRIT$) and the additional authentication as privileged ($PRIV$). 
To check for different levels of privilege (Requirement R\ref{req:priv}), we assume the presence of a \textit{privileged}  security level ($PRIV$) that is higher than (i.e., implies) normal authentication ($AUTH$). We further assume sensitive, \textit{critical} resources ($CRIT$), that need that higher security level (i.e., not everyone that is just authenticated can access these resources).
We note that from a formal perspective, $PROT$ and $CRIT$ are unrelated. The higher protection level of $CRIT$ is purely semantic. It stems from the requirement of a $PRIV$ which is hierarchically superior to $AUTH$, which is the sole requirement for $PROT$.
\Property{
\label{prop:priv}
$\square(({PRIV}\rightarrow{AUTH}) \land (\neg{PRIV}\land{CRIT}\rightarrow\neg{ACCESSOK}))$
}

\subsubsection{Key Validity}
\label{sec:check:rules:key}
A property for using a valid key (Requirement R\ref{req:key}) is straightforward, since it only has to define that an invalid key must not be accepted in the entire system.
\Property{
\label{prop:key}
$\square(\neg{INVKEYOK})$
}
Depending on the protocol, we use propositions like $WRONGKEYOK$ or $OLDKEYOK$ that would both be seen as invalid keys.

\subsection{Checking the Properties}
\label{sec:check:check}
Once we have a Rebeca model according to a template (Definition \ref{def:template}) and respective properties, the Rebeca Model Checker (RMC) can check the model for these properties.  Using a protocol-specific CPM, we define the propositions that constitute the properties in the context of a specific protocol. In our running example (see Figure \ref{fig:ExampleAMMtau}), we defined the necessary propositions for Property P\ref{prop:auth} only (for the sake of simplicity), which holds for this example. If we alter the transition for $\sigma=READ$ in $q=s0$ from $\omega=ERR$ to $\omega=OK$, then respective implicit ($\tau$) state gets the proposition $ACCESSOK$, but not $AUTH$ and the property will be violated. Semantically the latter case would mean a successful access without authentication.

\section{Testing}
\label{sec:test}
With a property violation discovered by the learning and checking procedure, the only thing left is to verify the finding on the actual system. Due to relying on conformance testing for equivalence queries, the automata learning process maintains a residual risk for an inaccurate model. We therefore test the trace of a found violation with the live system to exclude false positives. If the actual system behavior matches the one predicted by the model, the violation is confirmed. Otherwise, the trace can be fed back to the learning system to refine the model. We implement this step using a custom Oracle for the automata learner, where we can input the found counterexample for a new learning iteration. As a by-product, we use this custom oracle to improve the learning performance with a priori protocol knowledge by using test traces that resemble the standard behavior of a protocol first.
The verification of found traces is trivial, as the same tool set from the learning can be used for concretizing the input and interfacing with the SUL. Therefore, this combination
can be seen as method for test case generation.
\begin{table*}[t!]
\caption{CPM for eMRTDs.}
\label{tab:propNFC}
\begin{threeparttable}
\begin{tabularx}{\textwidth}{C|C|C}
\multicolumn{3}{X}{ }\\
\hline
\textbf{Proposition} & \textbf{Input} & \textbf{Output}\\
\hline
\multicolumn{3}{c}{Gains ($C_g$)}\\
\hline
AUTH	&	BAC	&	9000\\ 
DF, PROT	&	DF*	&	9000\\ 
EF	&	EF*	&	9000\\ 
CRIT	&	EF\_DG2, EF\_DG3
	&	9000\\ 
PRIV\tnote{a} & TA & 9000\\
\hline
\multicolumn{3}{c}{Losses ($C_l$)}\\
\hline
EF, AUTH, PRIV, CRIT	&	DF	&	9000\\ 
AUTH, PRIV	&	EF*, *BIN, *REC	&	6*\\ 
CRIT	&	EF*	&	9000\\ 
\hline
\multicolumn{3}{c}{Implicit State Propositions ($C_\tau$)}\\
\hline
UACCESSOK, ACCESSOK & SEL\_EF* & 9000\\
SSELEFOK, SACCESSOK, ACCESSOK & SSEL\_EF* & 9000\\
UREADOK, READOK	&	RD\_BIN	&	9000\\ 
SREADOK, READOK	&	SRD\_BIN	&	9000\\ 
SSELEFOK	&	SSEL\_EF*	&	9000\\ 
INVKEYOK, WRONGKEYOK & WS* & 9000\\
INVKEYOK, OLDKEYOK & OS* & 9000\\
\hline
\end{tabularx}
\begin{tablenotes}
\small
\item[a] Since, we do not have a Terminal Authentication implementation, this condition is hypothetical (will not be triggered).
\end{tablenotes}
\end{threeparttable}
\end{table*}

\section{Evaluation}
\label{sec:eval}
To show the practical usability and versatility of the described process, we evaluate it on two devices (an eMRTD and an automotive control unit), each speaking one of the communication protocols described earlier in this paper (NFC and UDS). Though the use cases are very different, in both cases the model checking is very similar. At the present abstraction level, we check generic security requirements using according LTL properties. The properties and CPMs are pre-defined (whereby the specification is fairly easy), while the annotated Mealy is automatically generated (utilizing a CPM) from a fully automatically learned Mealy machine. Thus, we can claim that the complete process of learning and checking a real-life eMRTD and UDS-based system is largely automated, with very little human interaction necessary.

\subsection{Electronically Machine-Readable Travel Document}

\subsubsection{Protocol Description -- Near Field Communication, Integrated Circuit Access, and Electronically Machine-Readable Travel Documents}
\label{sec:pre:NFC}
Near Field Communication (NFC) is a communication standard \cite{internationalorganizationforstandardizationCardsSecurityDevices2018,internationalorganizationforstandardizationCardsSecurityDevices2018a} that defines wireless communication for small, powerless embedded systems with low data rates. It defines messages for data transmission (information or I blocks), signaling (supervisory or S blocks), and acknowledgments (receive-ready or R blocks) and the modalities operate communications. The main applications are RFID-tags and wireless chip card access. For the latter, it uses the structures of ISO/IEC 7816-4 \cite{internationalorganizationforstandardizationIdentificationCardsIntegrated2020}. This standard logically segments circuits into applications (comparable to directories in a file system) that can be accessed via \textit{Dedicated Files (DFs)} and contains \textit{Elementary Files (EFs)} as data stores. The standard defines commands to access (\textit{SELECT}) data and applications, as well as manipulate data (\textit{READ, WRITE, UPDATE}, etc.).
It also defines \textit{GETCHALLENGE} and \textit{AUTHENTICATE} commands to implement challenge-response-based authentication mechanisms that protect sensitive data. Usually, an authentication procedure yields a session key, that is subsequently used to encrypt the access to the protected parts of the system. Every command induces a response that contains an unencrypted two-bytes status code (even if the data itself is encrypted). This code is either \textit{9000} for positive results or \textit{6XXX} for various error codes.

\begin{figure*}%[ht!]
	    \centering\includegraphics[width=\linewidth]{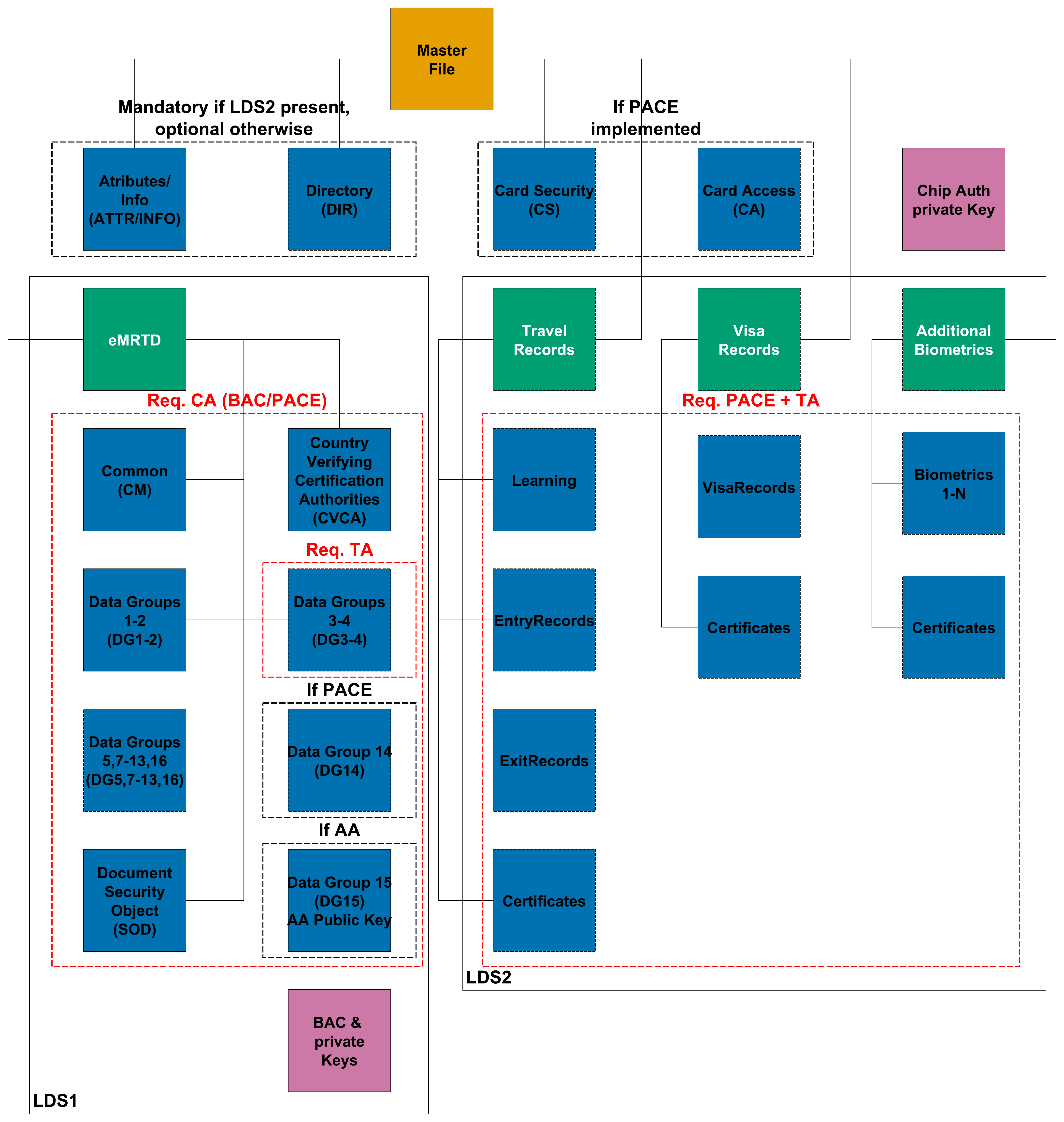}
\caption[width=\linewidth]{Logical Data Structure of Machine Readable Travel Documents from 
~\cite{marksteinerAutomatedPassportControl2024}. This shows the applications and files on an eMRTD chip. Amber is the master file (MF), Cyan are dedicated files (DF), Blue are Elementary Files (EF), and Green are key files. Solid frames means mandatory files, dashed ones optional files. Solid boxes donate the LDS contexts, dashed black boxes requirements, and dashed red boxes necessary authentication.
}
\label{fig:MRTD}
\end{figure*}

\label{sec:pre:emrtd}
NFC is also the main protocol to access Electronically Machine-Readable Travel Documents (eMRTDs). These provide a logical data structure to store travel document (e.g., passport) data on chips. The International Civil Aviation Organization (ICAO) standardizes this format in their  Doc 9303 part 10 standard~\cite{internationalcivilaviationorganizationMachineReadableTravel2021}. It defines two Dedicated File (DF) areas \textit{LDS1} (containing the eMRTD application) and \textit{LDS2} (containing  travel records, visa records, and additional biometrics). The eMRTD contains the Common (CM), the Country Verifying Certification Authorities (CVCA), and the Document Security Object (SOD) Elementary Files (EFs), as well as EFs for 16 data groups that contain various types of data, like personal data, document data, and biometrics (fingerprint, iris). The latter are more sensitive and therefore additionally protected, while the rest is protected by either the older  Basic Access Control (BAC) or the newer PACE (Password Authenticated Connection Establishment). BAC generates a key based on some cryptographic operations with  the passport number, expiration date, and the owner's date of birth. PACE uses a password to encrypt a nonce, which then is the base for a Diffie-Hellman-Merkle key exchange that creates a session key.
Besides there are some conditional files outside of both LDS, namely Attributes/Info (ATTR/INFO), Directory (DIR), Card Access (CA), and Card Security (CS). Figure \ref{fig:MRTD} shows an overview of the ICAO eMRTD schema.

\subsubsection{Checking eMRTDs}
\label{sec:check:rules:NFC}
For eMRTDs, we define $PROT$ as $DF$ input  in $C_g$ of the CPM for eMRTDs (Table \ref{tab:propNFC}), since the ICAO standard defines the applications to be in a protected zone. We only set DF for selecting LDS1, which corresponds to the eMRTD application. Technically, DF would also be set when selecting other applications (i.e., LDS2). However, we then would design the CPM to distinguish between DFs. This is not necessary in our case, since we only use one. $AUTH$ is gained by performing a \textit{basic authentication (BAC)} operation. As $ACCESSOK$ we define a successfully performed file selection in $C_\tau$. This means that for P\ref{prop:auth} a BAC must be performed before selecting any file inside an application.

For $UREADOK$, we define an unsecured \textit{read binary} operation. This means that for P\ref{prop:conf}, only secured \textit{read binary} operations are allowed.

We define $CRIT$ as a successful selection of data groups 2 or 3 in $C_\tau$. These contain biometric data, which is why the ICAO standard requires and additional (particularly terminal) authentication. Since, we do not have an additional authentication method implemented, $PRIV$ is set to false ($\bot$). This means that P\ref{prop:priv} only holds when these sensitive data groups cannot be selected in the complete model.

Lastly, we define $INVKEYOK$ for P\ref{prop:key} as \textit{any} successful secured operation that has been carried out using an old key or an all-zero key in $C_\tau$. We realize these by dedicated input symbols starting with WS (wrong key secured) and OS (old key secured).

    We also use a couple of complimentary properties for checking eMRTDs.
SecureRead:  $\square ( \neg  (SREADOK  \land   \neg (DF  \land  AUTH  \land  EF)))$,
which reads as a secure read operation (SREADOK) can only be successful if authenticated and a protected resource is selected ($DF \land EF$). This also creates the need for a normal read operation not to work on protected resources:
PlainRead:  $\square ( \neg  (UREADOK  \land  ( \neg EF  \lor  DF)))$,
which reads as a successful plain read (READOK) cannot happen without a selected resource outside of LDS1 ($\neg EF$$\lor DF$). We can further specify a secure select process with the following property:
SecureReadFollowsSecureSelect:  $( (\neg SREADOK)\ \mathcal{U}\ \allowbreak\text{SSELEFOK}) \allowbreak \lor   \square ( \neg SREADOK)$. This means that a secured read operation may only follow after a secured selection operation has occurred, independent of the EF/DF properties in the AMM. The reason is to prevent errors like that the EF proposition has been obtained through an unsecured select EF. Table \ref{tab:propNFC} gives an overview of the CPM.

\subsubsection{Results}
\label{sec:eval:NFC}
We scrutinize two different Austrian passports, an older, expired one and a current one. The model was learned via LearnLib using the TTT algorithm (with binary closure) and a minimum input trace length of 40 symbols, a maximum of 50, and 150 random walk conformance tests as equivalence oracle. This learning process is described in~\cite{marksteinerAutomatedPassportControl2024,marksteiner_learning_2025}. The checking process is two-fold: we once have to compile the generated model, which takes about 15 seconds, while the checking process itself takes around 4.5 seconds for the full state space (upper bound, which occurs if no violation has been found) and produced a space of 395 explored states and 956 transitions.
 \begin{figure*}[ht!]%[ht!]
	    \includegraphics[width=\linewidth]{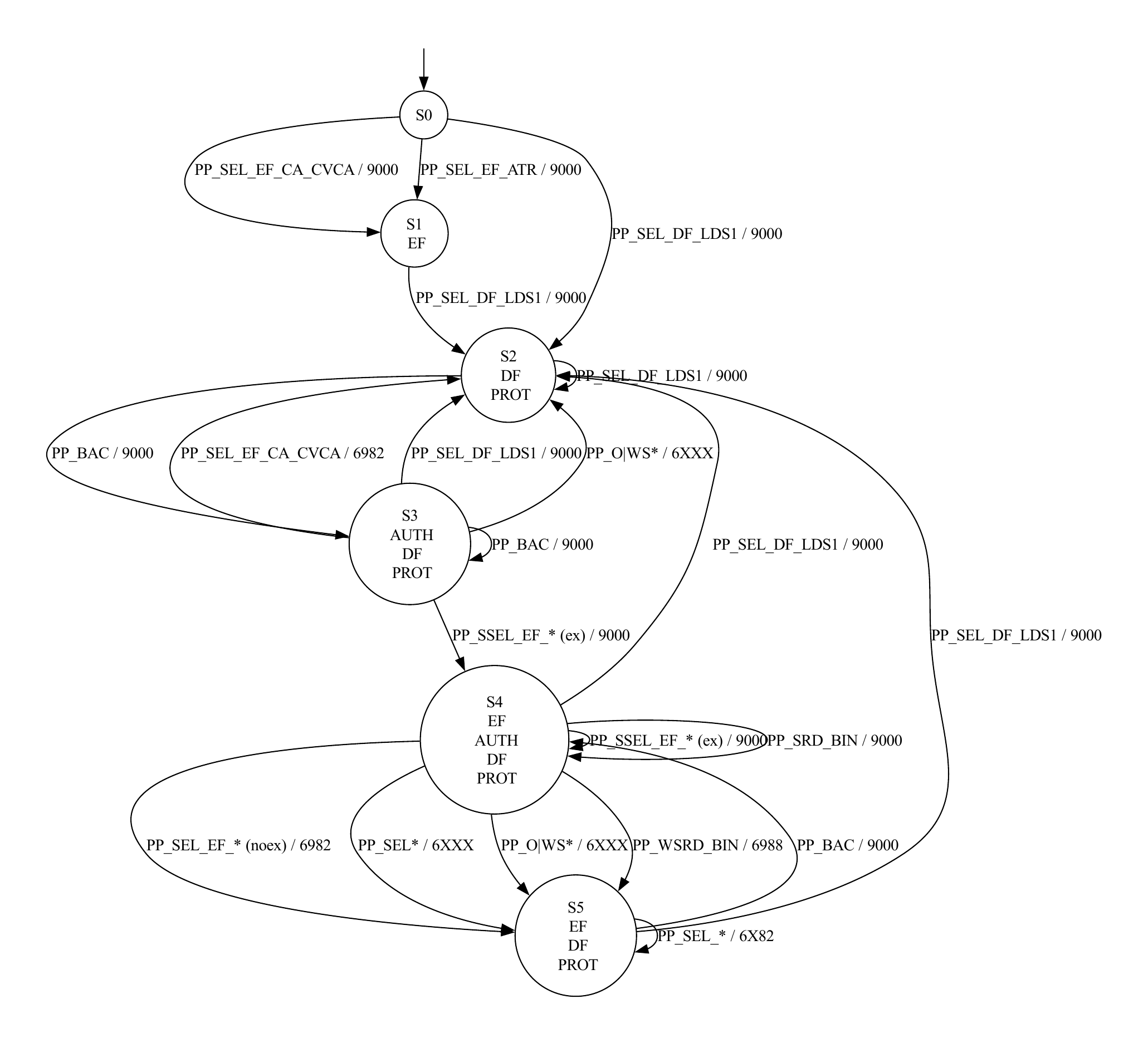}
\caption{Simplified version of the learned eMRDT's AMM. Self-loops that do not add to the understanding have been removed for readability. PP\_ is a general prefix for passport inputs. (S)SEL\_EF* (ex) and (noex) denotes all elementary file selections of existing and non-existing files. O|WS* denote secure commands that are encrypted with a wrong or old key. Note that with our CPM definition for eMRTDs, the protected attribute (PROT) is equal to a selected application (DF).}
\label{fig:NFCModel}
\end{figure*}
The result was a 6 state Mealy machine which we annotated the Mealy machine to an AMM using the method described above. The AMM (see Figure \ref{fig:NFCModel}) shows how files are selected (EF) inside and outside of the eMRTD application (DF/PROT) and the authentication inside the application (AUTH). It is noteworthy that a successful file read in this diagram only occurs via SRD\_BIN in state S4, which is a visual evidence that P\ref{prop:auth} and P\ref{prop:conf} hold. The higher privileged data groups where not selectable, since we do not have higher privileges (P\ref{prop:priv} holds). Old and wrong key commands result in de-authentication instead of a success, as can be seen with the transitions of PP\_O|WS (denoting old/wrong key operations) with an error in states S3 (falling back to s2) and S4 (falling back to S5), showing that P\ref{prop:key} holds. 
For the automated formal analysis of these properties through model checking, we automatically translated the AMM into Rebeca code (see the Listings  in \ref{sec:AppA} for parts of the model code for the environment, the system agent, and the property file respectively). 
Using the CPM in Table \ref{tab:propNFC} and properties as described in Section \ref{sec:check:rules:NFC}, we were able to verify that described security properties hold the scrutinized systems. The testing step is therefore not applicable, since no property violation was present. This means that we could verify in an automated pipeline, that Austrian passports implement authentication, confidentiality, privilege levels, and key validity.  Furthermore, we also successfully verified the additionally defined properties for passports, e.g. that a successful secure read may only occur after a successful secure select.

\subsection{Automotive Electronic Control Unit}

\subsubsection{Protocol Description -- Unified Diagnostic Services}
\label{sec:pre:UDS}
Unified Diagnostic Services (UDS) is an ISO standard that specifies services for diagnostic testers to control diagnostic (and further) functions on an on-vehicle Electronic Control Unit (ECU)~\cite{internationalorganizationforstandardizationRoadVehiclesUnified2020}. Therefore, it defines a variety of services, whereby the concrete elaboration mostly lies in the hands of an implementer. The protocol operates on layer 7 of the OSI reference model. Therefore, it relies on an underlying communication protocol. Although it can use multiple protocols, the most proliferated is the Controller Area Network (CAN) protocol. As a client-server-based protocol it defines requests and responses. The requests contain a service ID, optionally followed by sub-service IDs. Table \ref{tab:UDS} shows an overview of some of these services and their identifiers.
\begin{table}[t!]
\caption{Some of the most widely used UDS services.}
\label{tab:UDS}
\begin{threeparttable}
\begin{tabularx}{\columnwidth}{>{\centering\arraybackslash}X |                                c}
\hline
\textbf{Service} & \textbf{Identifier}\\
\hline
Diagnostic Session Control & 0x10\\
ECUReset & 0x11\\
ClearDiagnosticInformation & 0x14\\
ReadDTCInformation & 0x19\\
ReadDataByIdentifier & 0x22\\
ReadMemoryByAddress & 0x23\\
ReadScalingDataByIdentifier & 0x24\\
SecurityAccess & 0x27\\
CommunicationControl & 0x28\\
ReadDataByPeriodicIdentifier & 0x2A\\
DynamicallyDefineDataIdentifier & 0x2C\\
WriteDataByIdentifier & 0x2E\\
InputOutputControlByIdentifier & 0x2F\\
RoutineControl & 0x31\\
RequestDownload & 0x34\\
RequestUpload & 0x35\\
TransferData & 0x36\\
RequestTransferExit & 0x37\\
RequestFileTransfer & 0x38\\
WriteMemoryByAddress & 0x3D\\
TesterPresent & 0x3E\\
AccessTimingParameter & 0x83\\
SecuredDataTransmission & 0x84\\
ControlDTCSetting & 0x85\\
ResponseOnEvent & 0x86\\
ResponseControl & 0x87\\

\hline
\end{tabularx}
\end{threeparttable}
\end{table}
We will use a set of these services as an input alphabet for learning the UDS automaton.

\subsubsection{Checking UDS}
\label{sec:check:rules:UDS}
Authentication in UDS works over \textit{SecurityAccess} and subsequent \textit{SecurityAccessWithKey}. We therefore define the $AUTH$ proposition after the second step has been successfully executed. 
To verify property P\ref{prop:auth}, we perform a reading operation on a protected resource. We choose to start the \textit{Check Programming Dependencies} (0xFF01) routine. This routine is standardized~\cite{internationalorganizationforstandardizationRoadVehiclesUnified2020} and does not alter the firmware. Yet, it is industry practice that this route is protected by a \textit{Security Access} routine as it is a preparatory action for flashing the system~\cite{visachFlashBootloaderOEM2024}. 
Therefore we set both $PROT$ and $ACCESSOK$ as a $c_\tau$ when starting this routine.
For P\ref{prop:conf}, set perform a direct memory read operation to set $UREADOK$. Unfortunately, we do not have the knowledge of which memory are actually protected (as a workaround defining the CheckASWBit routine could also be defined as $PROT$). For P\ref{prop:priv}, we define the \textit{RequestDownload} routine as $CRIT$, as it allows to actually flash the device, while $PRIV$ is defined as \textit{SecurityAcccess} with a higher level. Lastly for P\ref{prop:key}, $INVKEYOK$ is defined as \textit{SecurityAcccess} with a wrong key (as an own input symbol).
Table \ref{tab:propUDS} shows the CPM for UDS.

\subsubsection{Results}
\label{sec:eval:UDS}
\begin{figure*}[t]%[ht!]
	    \includegraphics[width=\linewidth]{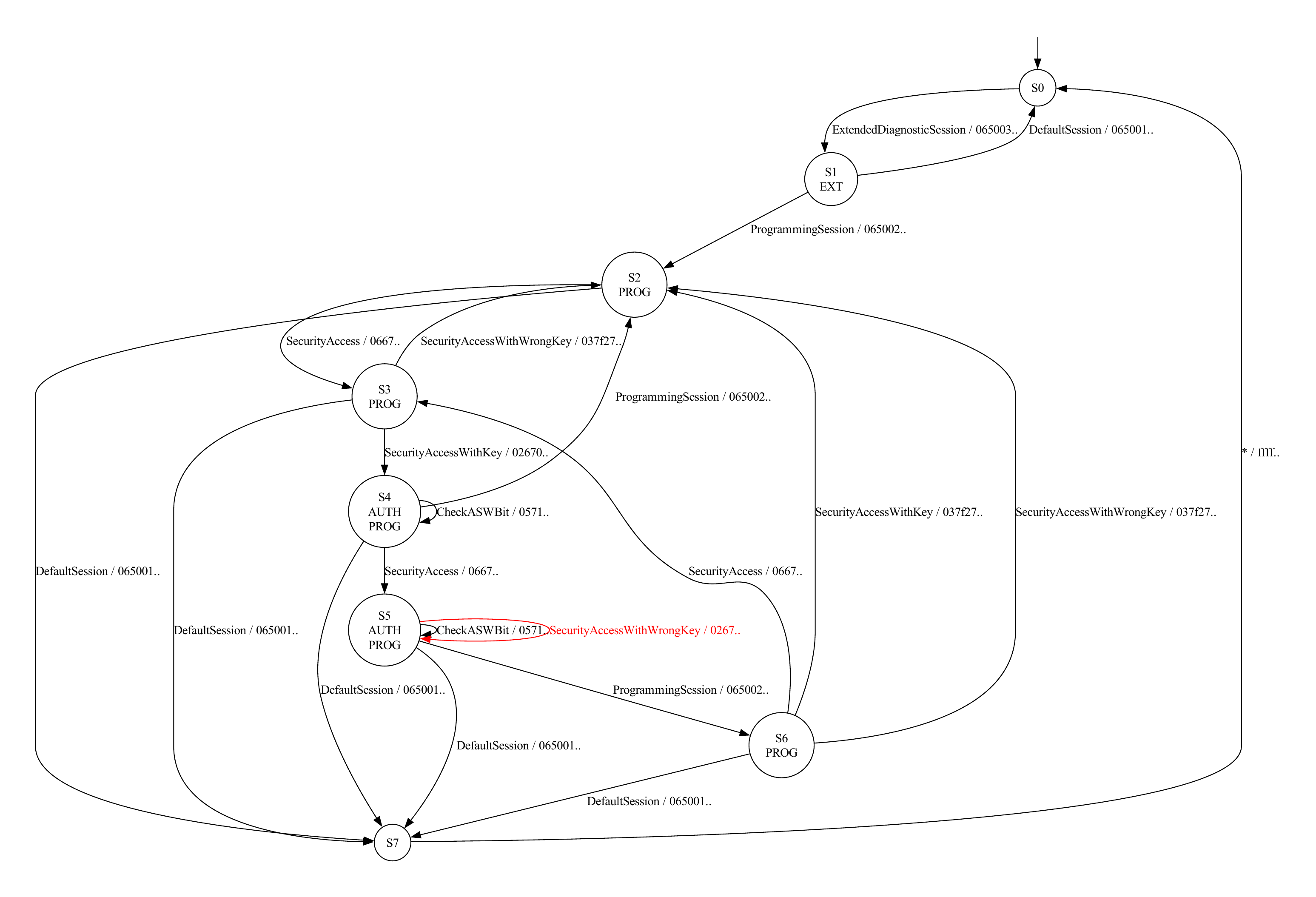}
\caption{Simplified AMM of the learned ECU model. Self-loops that do not add to the understanding have been removed and the output truncated for readability. The star (*) denotes any other input not explicitly stated. The transition in red shows the violation of security requirement R\ref{req:key} and property P\ref{prop:key}.}
\label{fig:UDSModel}
\end{figure*}

Our example use case for UDS is an automotive electronic control unit (ECU) from a major European Tier-1 supplier running in a vehicle from a Chinese car manufacturer. The model was learned via LearnLib using the TTT algorithm with a minimum input trace length of 20 symbols, a maximum of 50, and 50 random walks as equivalence oracle. We describe the learning of this model in~\cite{ebrahimiSystematicApproachAutomotive2023}.
We received an 8 state Mealy machine, which was annotated to an AMM
(see Figure \ref{fig:UDSModel}) using the CPM in Table \ref{tab:propUDS}) of this ECU.
 \begin{table*}[t!]
\caption{CPM for UDS (SA means SecurityAccess).}
\label{tab:propUDS}
\begin{threeparttable}
\begin{tabularx}{\textwidth}{C|C|C}
\multicolumn{3}{X}{ }\\
\hline
\textbf{Proposition} & \textbf{Input} & \textbf{Output}\\
\hline
\multicolumn{3}{c}{Gains ($C_g$)}\\
\hline
AUTH & SAWithKey & 67 \\
EXT & Extended & 5003 \\
PROG & Programming & 5002 \\
PRIV\tnote{a} & HLSAWithKey & 67 \\
\hline
\multicolumn{3}{c}{Losses ($C_l$)}\\
\hline
EXT,PROG & Default & 5001 \\
EXT & Programming & 5002 \\
PROG & Extended & 5003 \\
AUTH & SA, SAwKey, SAwWrongKey & 7f \\
AUTH & Session & 50 \\
\hline
\multicolumn{3}{c}{Implicit State Propositions ($C_\tau$)}\\
\hline
INVKEYOK, WRONGKEYOK & SAwWrongKey & 67 \\
PROT & CheckASWBit & 71 \\
ACCESSOK & CheckASWBit & 71 \\
UACCESSOK & CheckASWBit & 71 \\
CRIT & RequestDownload & 74 \\
UREADOK & Read* & 62 \\
\hline
\end{tabularx}
\begin{tablenotes}
\small
\item[a] Since, we do not have higher level security access, this condition is hypothetical (will not be triggered).
\end{tablenotes}
\end{threeparttable}
\end{table*}
Automatically translating the AMM into Rebeca (see the Listings  in \ref{sec:AppB} for parts of the model code for the environment, the system agent, and the property file respectively), 
we subsequently use the properties from Section \ref{sec:check:rules} to check the model for security properties, along with the CPM mentioned above. 
Like eMRTDs, we required a one-time effort of approx. 15 seconds to compile the generated model. Because of the smaller input alphabet, checking the ECU is a bit less complex than MRTDs taking around 4.2 seconds per property and creating a smaller state space with 131 states and 228 transitions for full model exploration.
We could verify P\ref{prop:auth}. For P\ref{prop:conf}, we lack the knowledge which memory areas are actually protected. However, it is well known that the CAN bus (over which UDS communication runs) is completely unencrypted, while principally critical (safety) data can be exchanged~\cite{wolf_secure_2006}. Therefore, R\ref{req:conf} is not met by the protocol itself, regardless of any implementation. To test P\ref{prop:priv}, we lack a higher layer security access. However, scrutinizing critical functions (\textit{RequestDownload}), we saw P\ref{prop:priv} to hold, as it were not possible to trigger with our current security level. P\ref{prop:key} was found to be violated. In the authenticated state S5 (see the transition in red in Figure \ref{fig:UDSModel}), the system accepts a wrong key\footnote{An issue known to us from previous works~\cite{ebrahimiSystematicApproachAutomotive2023}.} with a success return code. According to the standard, wrong key should not be acknowledged, even though a wrong key does not mean the security level must be locked (de-authentication)~\cite{international_organization_for_standardization_road_2020}. For the testing step, the learned input trace of S5 and a wrong key input can be applied to the system using the automata learning adapter class. We omitted this step, since this is a known issue on this device, so we can be sure this is not a false positive.
As a summary, we can state that the examined ECU only fulfills two out of four tested security requirements.

\section{Related Work}\label{sec:rel}
This work is based on previous work on model learning and  model checking~\cite{ebrahimiSystematicApproachAutomotive2023,marksteiner_learning_2025,marksteinerAutomatedPassportControl2024,marksteinerBlackboxProtocolTesting2025}. All of these papers have a different focus than the current work and do not contain CPMs, AMMs, or a Rebeca template. While \cite{ebrahimiSystematicApproachAutomotive2023} concentrates on a conceptual process and provides technical solutions for the learning part only, particularly we took our automotive ECU case study's model from this paper. Our work in~\cite{marksteiner_learning_2025} provides an extensive description and case studies for learning NFC and BLE systems. The other two approaches are based on equivalence checking and need a manually crafted specification model, which this paper does not need. The paper in \cite{marksteinerAutomatedPassportControl2024} provides an approach to compare a learned model with a partially defined specification Mealy machine (in Aldebaran format) and additional hard-coded checks in a Java program, it does not contain the use of a state-of-the-art model checker. We used the model learned in this paper for our eMRTD case study. The contribution of \cite{marksteinerBlackboxProtocolTesting2025} on the other hand contains an approach to have a secure specification model for equivalence checking. It consists of modeling a specification in Rebeca and model-check it for security properties. We then convert the state space diagram of the verified model into a Mealy machine that then serves for equivalence checking with learned implementation models. 
Peled et al.~\cite{peled_black_1999} have provided a very influential paper regarding black-box testing that combines automata learning with model checking. 
They learn a model (using Angluin's algorithm) create a cross product with a B{\"u}chi automaton created from the negate of an LTL property to check. If the accepted language is empty, the property is satisfied. Otherwise, the contained (non-empty) words it serve as a counterexamples that will be tested against a true system model. If these tests yield the same results, the counterexample poses a property violation. Otherwise, it serves as a counterexample for refining the model. They propose both an off-line and on-the-fly approach. The former uses a fully learned (i.e., a full run-through of Angluin's algorithm) model that is checked. The latter incorporates the checking into the learning process and stops it, if a violation is found, potentially greatly reducing the runtime.
Groce et al.~\cite{groce_adaptive_2002} advance the former approach by integrating the model checking deeper in the learning algorithm for an efficiency boost. 
Shijubo et al~\cite{shijubo_efficient_2021} expand Peled's approach by introducing strengthened specifications, that tend to find counterexamples that earlier detect divergence between the learned and the actual model, boosting the learning process' efficiency.
We refrain from using on-the-fly checking because we want to fully investigate the model for any violation of a defined set of properties, and therefore not stop at the first found violation,
as the purpose is to investigate already implemented real-world systems for all incorrect behavior. 
While the Rebeca model checker also stops with a counterexample when a property violation is found, we can, in contrast to on-the-fly-checking, check the other properties as well on the complete model and therefore determine if the system violates multiple properties. Furthermore, many properties do not actually require a modal operator -- e.g., Property~\ref{prop:auth} uses a single $\square$ operator, which means it should hold globally and a further distinction is not necessary. We can check these with a simple script running over a Rebeca-generated full state space, enabling us to find all occurrences of a violation.
While our work is based on a similar idea as the off-line variant, we extended the approach by annotating Mealy machines with atomic propositions creating AMMs that we turned into Rebeca code for model checking.
That way, we implemented all the phases for going through learning, checking, and testing real-world examples. In our case, however, we automated the complete process by combining (a) the learner(s) (using LearnLib~\cite{isberner_opensource_2015}), (b) an automated way of annotating Mealy machines with propositions, creating AMMs  and translating it into a checkable model, and (c) a model checker translates this model and (the negation) LTL properties to B{\"u}chi automata, creates a product, and check its language for emptiness~\cite{khamespanahAfraEclipseBasedTool2023} with generic properties.
This improves both the usability of the approach and its applicability to real-world systems.

Neider and Jansen used a symbolic approach to learn and model-check DFAs (but not Mealy or More machines)~\cite{neider_regular_2013}. We extend this approach by automating the missing link between learning and model checking with CPMs. We note that our approach of automating cannot be applied to DFAs as used by them, since it the proposition annotation mechanism based on CPMs relies on system output behaviors that is not visible in DFAs.
Similarly, Fiterau-Brostean et al.~\cite{fiterau-brostean_combining_2016,fiterau-brostean_model_2017} also used a symbolic approach (using the NuSMV model checker). 
Translating the learned model to Rebeca, our approach in contrast provides more advanced possibilities to manipulate the model, including re-introducing abstracted (e.g. non-deterministic) behavior, as they also stated in their work that ironing out timing issues were a major engineering problem. Altering the Rebeca template, it is trivial to re-introduce timeouts (non-deterministically) to check a protocol's behavior under such conditions. This allows for more advanced checking possibilities. Furthermore, our approach using CPMs does not only allow for a more explicit and comprehensible annotation of states with propositions, but also using generic properties and a clean separation between state propositions (visible in the annotated Mealy machine) and pure transition outputs (in $C_\tau$ and $\tau$ states).
This allows for a separating the structure (in the Rebeca template and the generic properties) and the content (the propositions defined by CPMs). The structure can be reused without changes for multiple use cases, while the content can be defined quickly and easily using CPMs.
For the principle conversion of LTS (which Mealy machines could be seen as) and Kripke structures there is a well-known standard procedure by De Nicola and Vaandrager~\cite{de_nicola_action_1990}, used by many scholarly works . However, our approach differs by utilizing both state and transition labels' information to answer very specific questions in the security context (e.g., about resource access and prior authentication) using implicit state propositions based on the state labels. Also, the proposed AMM structure is more convenient to transfer into a Rebeca model, which is beneficial, as stated above. Furthermore, we argue that the CPM approach provides an easy access for model checking, as these are easy to create and the rest of the process is automated.
Other approaches~\cite{schuts_refactoring_2016,tapplerModelBasedTestingIoT2017,marksteiner_learning_2025} combine automata learning with bisimulation-based equivalence checking, which  need a specification model to compare the learned automaton with. Our approach  needs generalized properties. We do not know of an approach that translates a learned model into a modeling language allowing for both direct property checking 
and manipulating the model to simulate scenarios.

\section{Conclusion}
\label{sec:conc}
In this paper, we presented an approach to combine automata learning with model checking for reactive systems.
We defined \textit{Context-based Proposition Maps (CPMs)} that provide an annotation mechanism to annotate Mealy machines inferred by automata learning with propositions. As a result, we received \textit{annotated Mealy machines (AMMs)} that combine attributes of Mealy machines and Kripke structures. We also formally defined these AMMs as an extension of classical Mealy machines combined with a set of properties and a (state) labeling function.
We subsequently translated the AMMs into Rebeca code in an automated process by formally defining a template using two actors: a receiver modeling the learned system and a sender modeling an external actor that interacts with the system. 
Starting from four high-level security requirements (authentication, confidentiality, privilege levels, and key validity), we defined generic LTL properties to check models for security. Protocol-specific CPMs provided the context to assure the generated Rebeca code provides all propositions that are used in these properties. This way, each property can be checked with regard to the protocol the examined system runs on. We  presented a case study with systems from different domains (two passports and an automotive control unit) speaking different protocols (NFC and UDS) to show the versatility of the approach on real-world off-the-shelf systems. We were able to verify the security properties for all requirements on the two different passports. On the automotive control unit, our verification process verified only two of them (authentication and privilege levels). The third requirement (confidentiality) is intrinsically not provided by the protocol. The last one (key validity) was not met (property violated) on the examined unit, which means we could falsify this requirement. These techniques applied in the presented case studies provide evidence for the usefulness of the approach, since it significantly reduces modeling and property design efforts through automation. Given a working adapter for learning a protocol, the only manual effort to be taken by the verifier is to define a corresponding CPM. As we outlined in this paper, this task is not particularly resource-intensive. 
CPMs are not only intentionally designed to minimize the required manual interaction, but also remove the necessity for expertise in formal methods, since engineers defining a CPM only need knowledge about the scrutinized protocol.  However, a comprehensive user experience study would go beyond of the scope of this work and pose a separate research contribution. 
As a summary, we provided security properties for an automated model checker (RMC) by deriving them from generic security requirements and assigning their propositions with CPMs. On the other hand we automatically derived checkable models by generating Mealy machines through automata learning and annotating them to AMMs using the same CPMs as for the properties.

\subsection{Limitations}
We note that our approach has some limitations. While we have a fully working tool chain from learning to checking, we stress that the scope of this work is the missing piece to convert learned Mealy machines to a form suitable for checking security by adding context-loaded propositions and not the learning or model checking itself. Also, the learning could be incomplete, which is subject of the underlying automata learning framework. We partly mitigate this shortcoming by treating found property violations as test cases. That means we feed the trace of a found violation into the tested system and render it an actual issue only if the system under test behaves as predicted. Otherwise, the trace poses a counterexample we can feed back to the learning algorithm to refine the Mealy machine. However, we can assure the checkable model we generate is faithful to the learned one. We built in a feedback loop that converts the explored state space from the model checking back to a Mealy machine and performs an equivalence check with the originally learned model. The correctness of the CPM on the other hand is out of scope, and fully depends on the manual definition -- plausibility checks or even verification against standards documents are possible, but subject to future work. Lastly, we state that the degree of generality of the approach is high but not total. While the security properties are general, the underlying propositions have to be contextualized with a hard connection to a protocol by the CPMs. However, this specialization is still broad, since one protocol CPM works with all implementations of that protocol.

\subsection{Outlook}
\label{sec:conc:out}
Further research directions to lift the approach to a larger scale lie in creating adapters for other different protocols and creating other generalized LTL properties to check different aspects of security. Additionally, we are also investigating utilizing an LLM to generate LTL properties from threat models to check their respective implementations' Mealy machines. We can also improve the learning process by using timing~\cite{vaandrager_learning_2021} and by extending the model with unknown inputs using input symbol mutation; if the unknown input triggers new behavior, we can dynamically extend the input alphabet (as described here~\cite{schogler_automata_2023}). This allows for a more holistic testing. Furthermore, we plan to extend the CPMs to be more expressive (e.g., allow for logic operators in the CPM conditions or add an operator to assign propositions to the initial state).

\section{Acknowledgement}
This work has received funding from the European Union under the Horizon Europe programme under grant agreement No. 101168438 (project INTACT) as well as from the Swiss State Secretariat for Education, Research and Innovation and UK Research and Innovation under grant No. 10126241.
We further acknowledge the support of the Swedish Knowledge Foundation via the industrial doctoral school RELIANT under grant No. 20220130. 
The authors want to thank the anonymous reviewers of the Computers \& Security Journal, as well as of the International Conference on Runtime Verification for their valuable input. We further want to thank Thomas Grandits for helping to formulate requirements and properties more understandable.
%%%%%
\bibliography{literature}

\appendix
\section{eMRTD Rebeca Code Listings}
\label{sec:AppA}
\begin{lstlisting}[label=lst:NFCsend,caption=Example of generated eMRTD environment actor code.,language=rebeca,escapechar=|]
reactiveclass ENVIRONMENT(3) {
	knownrebecs {
		SYSTEM system;
	}
    [..]
	ENVIRONMENT() {
	  self.start(); 
	}
	void start() {
		system.req();
	}
	msgsrv req_6986(){
		system.req();
	}
    [..]
   	msgsrv req_9000(){
		system.req();
	}
    [..]
}
\end{lstlisting}
\begin{lstlisting}[label=lst:NFCrec,caption=Example of generated eMRTD system actor code.,language=rebeca,escapechar=|]
reactiveclass SYSTEM(3) {
	knownrebecs {
		ENVIRONMENT environment;
	}
	statevars {
		boolean ef, df, prot, auth;
		int state;
		boolean error;
		boolean ureadok, sreadok, readok, sselefok, selefok, accessok, uaccessok, saccessok, invkeyok, wrongkeyok, oldkeyok;
	}
    msgsrv req() {
       error=false;
       ureadok=false;
       sreadok=false;
       readok=false;
       sselefok=false;
       selefok=false;
       accessok=false;
       [..]
       int data =?(0,1,2,[..],56);
       switch(data) {
          case 0: system.pp_sel_ef_ca_cvca(); break; 
          case 1: system.pp_sel_df_lds1(); break; [..]
          case 22: system.pp_rd_bin(); break; [..]
          case 28: system.pp_bac(); break; [..]
          case 32: system.pp_ssel_ef_dg1(); break;[..]
       } 
    }
    [..] 
    msgsrv pp_rd_bin(){
       if(state==0) {
          state=0; environment.req_6986();
       } else
       if(state==1) {
          state=1; environment.req_6986();
       } else
       if(state==2) {
          auth=false; state=1; environment.req_6986();
       } else
       if(state==3) {
          state=3; environment.req_9000(22);
       } else
       if(state==4) {
          auth=false; state=5; environment.req_6982();
       } else
       if(state==5) {
          state=5; environment.req_6982();
       }
    }
    [..]
    msgsrv pp_sel_df_lds1(){
       if(state==0) {
          df=true; prot=true; state=1; environment.req_9000(1);
       } else
       if(state==1) {
          state=1; environment.req_9000(1);
       } else
       if(state==2) {
          auth=false; state=1; environment.req_9000(1);
       } else
       if(state==3) {
          df=true; prot=true; ef=false; state=1; environment.req_9000(1);
       } else
       if(state==4) {
          auth=false; ef=false; state=1; environment.req_9000(1);
       } else
       if(state==5) {
          ef=false; state=1; environment.req_9000(1);
       }
    }
    [..]
    msgsrv pp_bac(){
       if(state==0) {
          state=0; environment.req_6985();
       } else
       if(state==1) {
          auth=true; state=2; environment.req_9000(28);
       } else
       if(state==2) {
          state=2; environment.req_9000(28);
       } else
       if(state==3) {
          state=3; environment.req_6985();
       } else
       if(state==4) {
          state=4; environment.req_9000(28);
       } else
       if(state==5) {
          auth=true; state=4; environment.req_9000(28);
       }
    }
    [..]
    msgsrv pp_ssel_ef_dg1(){
       if(state==0) {
          state=0; environment.req_6988();
       } else
       if(state==1) {
          state=1; environment.req_6988();
       } else
       if(state==2) {
          ef=true; state=4; environment.req_9000(32);
       } else
       if(state==3) {
          state=3; environment.req_6988();
       } else
       if(state==4) {
          state=4; environment.req_9000(32);
       } else
       if(state==5) {
          state=5; environment.req_6988();
       }
    }
    [..]
\end{lstlisting}
\begin{lstlisting}[label=lst:NFCprop,caption=Example of the eMRTD property file.,language=rebeca,escapechar=|]
property {
	define {
		ERROR = system.error;
		EF = system.ef;
		DF = system.df;
		PROT = system.prot;
		AUTH = system.auth;
		S0 = (system.state ==0);
		S1 = (system.state ==1);
		S2 = (system.state ==2);
		S3 = (system.state ==3);
		S4 = (system.state ==4);
		S5 = (system.state ==5);
		UREADOK = system.ureadok;
		SREADOK = system.sreadok;
		READOK = system.readok;
		SSELEFOK = system.sselefok;
		SELEFOK = system.selefok;
		ACCESSOK = system.accessok;
		UACCESSOK = system.uaccessok;
		SACCESSOK = system.saccessok;
		INVKEYOK = system.invkeyok;
		WRONGKEYOK = system.wrongkeyok;
		OLDKEYOK = system.oldkeyok;
		CRIT = false;
		PRIV = false;
	}
	Assertion{
		Assertion0: (!ERROR);
	}
	LTL{
		P1:G( (!(!AUTH &&PROT)) || !ACCESSOK);
		P2:G( (!PROT) || (!UREADOK));
		P3:G( (!PRIV || AUTH) && ((!(!PRIV &&CRIT)) || !ACCESSOK));
		P4:G(!INVKEYOK);
		PlainRead: G(! (UREADOK && (!EF || DF)));
		SecureRead: G(! (SREADOK && !(DF && AUTH && EF)));
		SecureSelect: G(! (SSELEFOK && !(DF && AUTH)));
		ReadFollowsSelect: U(!READOK,EF) || G(!READOK);
		SecureReadFollowsSecureSelect: U(!SREADOK,SSELEFOK) || G(!SREADOK); 
	}
}
\end{lstlisting}

\section{UDS Rebeca Code Listings}
\label{sec:AppB}
\begin{lstlisting}[label=lst:UDSsend,caption=Example of generated eMRTD environment actor code.,language=rebeca,escapechar=|]
reactiveclass ENVIRONMENT(3) {
	knownrebecs {
		SYSTEM system;
	}
    [..]
	ENVIRONMENT() {
	  self.start(); 
	}
	void start() {
		system.req();
	}
	msgsrv req_037f2700000000aa(){
		system.req();
	}
    [..]
   	msgsrv req_065002003200c8aa(){
		system.req();
	}
    [..]
}
\end{lstlisting}
\begin{lstlisting}[label=lst:UDSrec,caption=Example of generated eMRTD system actor code.,language=rebeca,escapechar=|]
reactiveclass SYSTEM(3) {
	knownrebecs {
		ENVIRONMENT environment;
	}
	statevars {
		boolean ext, auth, prog;
		int state;
		boolean error;
		boolean invkeyok, wrongkeyok, prot, accessok, uaccessok, crit, ureadok;
	}
    msgsrv req() {
        error=false;
		invkeyok=false;
		wrongkeyok=false;
		prot=false;
		accessok=false;
		uaccessok=false;
		crit=false;
        ureadok=false;
       [..]
       int data =?(0,1,2,[..],56);
       switch(data) {
          case 0: self.defaultsession(); break;
          case 1: self.programmingsession(); break;
          case 2: self.extendeddiagnosticsession(); break;
          case 3: self.communicationcontrol(); break;
          case 4: self.securityaccess(); break;
          case 5: self.securityaccesswithkey(); break;
          case 6: self.securityaccesswithwrongkey(); break;
          case 7: self.testerpresent(); break;
          case 8: self.checkaswbit(); break;
          sdefault: self.ERR();
       } 
    }
    [..] 

        }
 }
    [..]
    msgsrv securityaccesswithwrongkey(){
        if(state==0) {
         state=0; environment.req_037f277faaaaaaaa();
        } else
        if(state==1) {
         state=1; environment.req_037f2712aaaaaaaa();
        } else
        if(state==2) {
         state=2; environment.req_037f2724aaaaaaaa();
        } else
        if(state==3) {
         state=0; environment.ffffffffffffffff();
        } else
        if(state==4) {
         state=2; environment.req_037f2735aaaaaaaa();
        } else
        if(state==5) {
         state=5; environment.req_037f2724aaaaaaaa();
        } else
        if(state==6) {
         state=6; invkeyok=true;wrongkeyok=true;environment.req_02670aaaaaaaaaaa();
        } else
        if(state==7) {
         state=2; environment.req_037f2735aaaaaaaa();
        }
 }
    [..]
\end{lstlisting}
\begin{lstlisting}[label=lst:UDSprop,caption=Example of the eMRTD property file.,language=rebeca,escapechar=|]
property {
	define {
		ERROR = system.error;
		EXT = system.ext;
		AUTH = system.auth;
		PROG = system.prog;
		S0 = (system.state ==0);
		S1 = (system.state ==1);
		S2 = (system.state ==2);
		S3 = (system.state ==3);
		S4 = (system.state ==4);
		S5 = (system.state ==5);
		S6 = (system.state ==6);
		S7 = (system.state ==7);
		INVKEYOK = system.invkeyok;
		WRONGKEYOK = system.wrongkeyok;
		PROT = system.prot;
		ACCESSOK = system.accessok;
		UACCESSOK = system.uaccessok;
		CRIT = system.crit;
		UREADOK = system.ureadok;
		PRIV=false;
	}
	Assertion{
		Assertion0: (!ERROR);
	}
	LTL{
		P1:G( (!(!AUTH &&PROT)) || !ACCESSOK);
		P2:G( (!PROT) || (!UREADOK));
		P3:G( (!PRIV || AUTH) && ((!(!PRIV &&CRIT)) || !ACCESSOK));
		P4:G(!INVKEYOK);
	}
}

\end{lstlisting}

\end{document}